\documentclass[prb, twocolumn, superscriptaddress, showpacs,amsmath,amssymb]{revtex4}
\usepackage{float}
\usepackage{graphicx}

\begin{document}

\title{Vertical distribution of nitrogen-vacancy centers in diamond formed by ion implantation and annealing}
\author{Charles Santori
\email{charles.santori@hp.com}}
\author{Paul E. Barclay}
\author{Kai-Mei C. Fu}
\author{Raymond G. Beausoleil}
\affiliation{Information and Quantum System Lab, Hewlett-Packard
Laboratories, 1501 Page Mill Road, MS1123, Palo Alto, California
94304, USA}

\begin{abstract}

Etching experiments were performed that reveal the vertical distribution of optically active nitrogen-vacancy (NV) centers in diamond created in close proximity to a surface through ion implantation and annealing.  The NV distribution depends strongly on the native nitrogen concentration, and spectral measurements of the neutral and negatively-charged NV peaks give evidence for electron depletion effects in lower-nitrogen material.  The results are important for potential quantum information and magnetometer devices where NV centers must be created in close proximity to a surface for coupling to optical structures.

\end{abstract}
\maketitle

\section{Introduction}

Nitrogen-vacancy (NV) centers located deep within a diamond lattice appear promising as solid-state spin qubits since they combine optical initialization and readout capabilities~\cite{ref:vanoort1988ods, ref:jelezko2004obs, ref:hanson2006rtm}, long electron spin coherence lifetimes (approaching $1\,\mathrm{ms}$ at room temperature~\cite{ref:gaebel2006rtc}), and the ability to control coupling to individual nuclear spins~\cite{ref:jelezko2004ocn, ref:childress2006cdc, ref:gurudevdutt2007qrb}.  While these capabilities were initially demonstrated in high-purity natural diamond samples, similar results have recently been achieved in high-purity CVD-grown diamond~\cite{ref:kennedy2003lct, ref:mizuochi2008csp}.  There is now much interest in fabricating spin-based devices in diamond, with potential applications in quantum communication~\cite{ref:childress2005ftq}, quantum computation~\cite{ref:nizovtsev2005qcb, ref:benjamin2006bgs}, and magnetometry~\cite{ref:taylor2008hsd, ref:degen2008smf, ref:balasubramanian2008nim, ref:maze2008nms}.  In all of these applications it is necessary or at least advantageous to couple NV centers to optical structures such as waveguides~\cite{ref:hiscocks2008dwf, ref:fu2008cnv} and microresonators~\cite{ref:park2006cqw, ref:greentree2006ccd, ref:tomljenovichanic2006dbp, ref:wang2007owg, ref:wang2007fct, ref:bayn2008uhq, ref:barclay2008cie}, to enable communication between distant qubits or to allow efficient extraction of emitted photons.  Therefore, a reliable method is needed to create NV centers with good spectral properties in close proximity ($< 100 \, \mathrm{nm}$) to a diamond surface.  In addition, the charge state of the NV center must be controlled; all of the results mentioned above were demonstrated using NV centers in their negatively-charged state, NV$^-$.

It has been well established that NV centers can be created by ion implantation or irradiation followed by annealing~\cite{ref:davies1976oso, ref:davies1992vrc, ref:mita1996cas, ref:martin1999gad, ref:waldermann2007cdc, ref:wee2007tpe}.  Ion implantation is used to create damage (carbon vacancies) in the diamond lattice, and it can also be used to introduce nitrogen impurities~\cite{ref:burchard2005gsc, ref:greentree2006ccd, ref:rabeau2006ils, ref:weis2008sad}.  The choices of particle and acceleration voltage determine the initial vertical distribution of vacancies.  Annealing is performed at temperatures (typically $600-1000^\circ \mathrm{C}$) where vacancies become mobile and can combine with nitrogen impurities to form NV centers.  Vacancies can also recombine with self-interstitials~\cite{ref:davies1992vrc, ref:allers1998air, ref:newton2002red} (implantation produces one interstitial carbon for each vacancy), can join together to form extended defects~\cite{ref:hounsome2006obc}, or can become trapped at dislocations or at the surface~\cite{ref:nelson1983dsi}.  While numerous annealing experiments have been performed, and vacancy migration over long distances in the horizontal direction has been observed at $1400^\circ \, \mathrm{C}$ annealing temperature~\cite{ref:gippius1999fcg}, to our knowledge none of these studies has yet clearly answered the question of how far vacancies diffuse vertically from the initially damaged layer at temperatures optimized to create NV centers.  This is of vital importance for positioning NV centers close to an optical structure.

Here, we present etching experiments that reveal the vertical distribution of optically active NV centers produced using ion implantation and annealing to convert native nitrogen into NV centers.  The results indicate that vacancies can diffuse several hundred nanometers to form NV centers at the annealing temperatures used.  Different behavior was observed for high-nitrogen ($\approx 100 \, \mathrm{ppm}$) and low-nitrogen ($\approx 1 \, \mathrm{ppm}$) diamond.  In low-nitrogen diamond, spectroscopy results show an unexpected dependence of the NV$^-$ signal on etch depth, and we interpret this behavior as an electronic depletion effect.

\section{Experiment}

\begin{figure}[ht]
\centering
\includegraphics[width=3.5in]{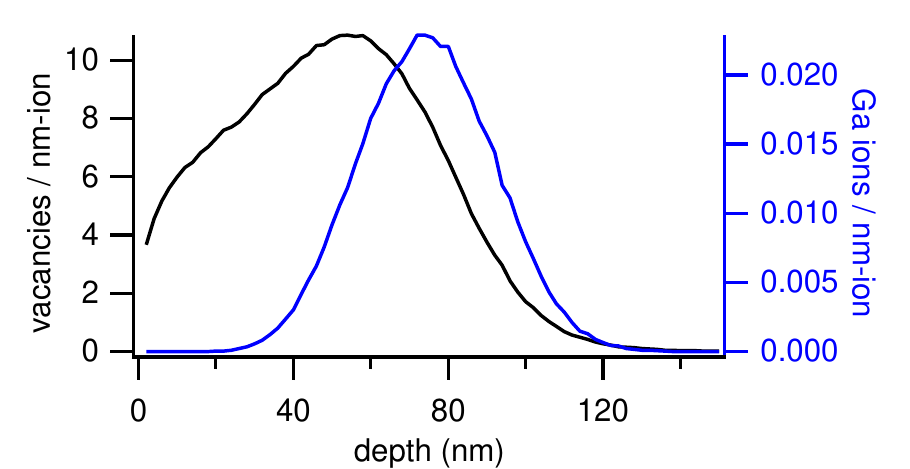}
\caption{Monte-Carlo simulations (SRIM) of the carbon vacancy (black) and Ga (blue) profiles produced by $200\,\mathrm{keV}$ Ga$^+$ ion implantation.}
\label{fig:implant}
\end{figure}

The first sample used in the experiments was a $3 \times 3 \times 0.5 \, \mathrm{mm}$, $(100)$-oriented CVD diamond (Element 6) with a specified nitrogen concentration below $1 \, \mathrm{ppm}$, showing fairly uniform NV concentration before processing.  The sample was implanted with $200\,\mathrm{keV}$ Ga ions (Core Systems) with a dose of $3 \times 10^{11} \, \mathrm{cm}^{-2}$.  A Monte-Carlo simulation using SRIM~\cite{ref:zeigler2008sri} predicts that this implantation energy produces vacancies to a depth of approximately $100\,\mathrm{nm}$, as shown in Fig.~\ref{fig:implant}.  During implantation, the CVD sample was covered with a TEM grid to allow measurement of the photoluminescence (PL) intensity contrast between implanted and non-implanted surfaces.  This technique was necessary because the CVD sample has a high background NV concentration relative to its nitrogen content.  After implantation, the sample was annealed at $925^\circ \, \mathrm{C}$ in an H$_2$/Ar forming gas for 3 hours.

The second sample was a $3\times 3 \times 0.5 \, \mathrm{mm}$, $(100)$ HPHT diamond (Sumitomo) with a specified nitrogen concentration of $30-100 \, \mathrm{ppm}$.  This sample has growth sectors with widely varying impurity concentrations.  In most sectors, the width of the NV$^-$ zero-phonon line (ZPL) measured at low temperature is broad ($\sim 1 \, \mathrm{nm}$), suggesting a high nitrogen concentration, and the NV luminescence achievable by implantation and annealing in these sectors is very high.  However, as we have previously observed in Sumitomo HPHT samples~\cite{ref:santori2006cpta}, there exist particular sectors, with boundaries aligned to the sample edges, that have properties consistent with much lower nitrogen concentration, including a narrow ZPL linewidth (as low as $\sim 0.02 \, \mathrm{nm}$) at low temperature and an inability to produce strong NV luminescence with high implantation doses.  The spectral properties of this low-nitrogen sector in the HPHT sample appear more similar to those of the CVD sample described above.  The HPHT sample was implanted with $200\,\mathrm{keV}$ Ga ions (Core Systems) with a dose of $3 \times 10^{12} \, \mathrm{cm}^{-2}$.  A TEM grid was not used on this sample, and was not required because the HPHT sample has a very low background NV concentration relative to its nitrogen content.  After implantation, the sample was annealed at $925^\circ \, \mathrm{C}$ in an H$_2$/Ar forming gas for 3 hours.
\begin{figure}[ht]
\centering
\includegraphics[width=3.0in]{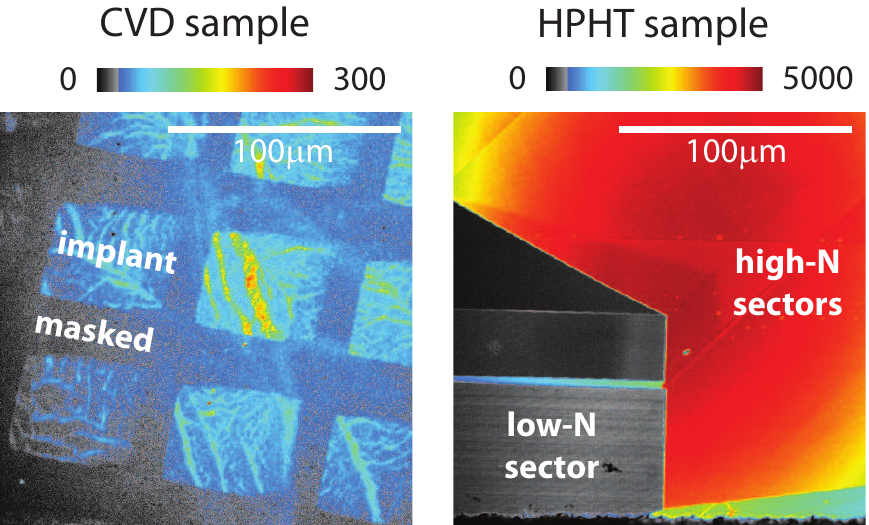}
\caption{Fluorescence images of the samples obtained by scanning confocal microscopy before etching: (a) Element 6 CVD sample, showing implantation grid pattern; (b) Sumitomo HPHT sample, showing growth sector boundaries.  Scan size: $178 \times 178 \, \mu\mathrm{m}$}
\label{fig:beforeimages}
\end{figure}

The samples were etched using an Oxford inductively coupled plasma reactive ion etching (ICP-RIE) system with an O$_2$/Ar recipe
\cite{ref:leech2001jms}.  Before etching, protective masks (SiO$_2$ or SiN) were patterned on portions of the samples to allow measurement of the etch depth and for observation of the PL contrast between etched and non-etched regions.  After each trial, the etch depth was measured using a profilometer.  Since the step height seen by the profilometer includes both the diamond etch depth and the mask thickness, for some trials we also measured the etch depth by electron microscopy.  We measured the optical properties of the NV centers using a scanning confocal microscopy setup.  For excitation, a $532\,\mathrm{nm}$ laser was focused through a 0.6 numerical-aperture microscope objective to a $\sim 0.5 \, \mu \mathrm{m}$ spot, and galvanometer mirrors or a piezo stage were used to scan the position of this spot on the sample.  The resulting PL was collected through the same objective and imaged onto a pinhole for spatial filtering, and detected using a cooled-CCD spectrometer or an avalanche photodiode with bandpass filters selecting wavelengths from approximately $647-804\,\mathrm{nm}$, for detection of the NV$^-$ phonon sidebands.  The depth of focus was such that when the signal from a thin sheet of NV centers was measured as a function of the height of the microscope objective, a symmetric peak was observed with a full width at half maximum of $3.8\,\mu\mathrm{m}$, corresponding to $9\,\mu\mathrm{m}$ in diamond.  The measurements were performed at room temperature.  For both samples, images of the PL intensity were obtained at the same locations after each trial.  For the HPHT sample, spectra were also obtained at multiple locations, and from these spectra the intensities of the NV$^0$ and NV$^-$ zero-phonon lines at $575\,\mathrm{nm}$ and $638\,\mathrm{nm}$, respectively, were individually determined.

Fig.~\ref{fig:beforeimages} shows PL images of the two samples before etching.  For the CVD sample, the implantation pattern is visible, combined with ribbon-like features that must be caused by a non-uniform nitrogen concentration present in the sample before processing.  For the HPHT sample, the image shows a sharp boundary between growth sectors with different nitrogen concentrations.  The region that appears dark is the low-nitrogen sector mentioned above.  From images similar to these, the total photoluminescence intensity in the $647-804\,\mathrm{nm}$ band was measured as a function of etch depth for the two samples.  The results, normalized by excitation power, are plotted in Fig.~\ref{fig:APDplots}.  For the CVD sample, the quantity plotted is the average intensity measured in the implanted squares minus the average intensity measured outside of the squares.  In this way we can separate the PL produced by NV centers created through ion implantation from the background NV PL.  For the HPHT sample, the initial NV concentration was very small, and only the total PL intensity is plotted in the figure, measured in both the high-nitrogen and low-nitrogen sectors.  Since the optical detection efficiency in our setup is approximately constant over $1 \, \mu\mathrm{m}$ of depth, the plotted quantities represent the total PL intensity from all of the remaining NV centers that were created through ion implantation and annealing.

Before presenting a detailed analysis, we note several important features of the data in Fig.~\ref{fig:APDplots}.  First, the behavior of the high-nitrogen sector of the HPHT sample for the first $300\,\mathrm{nm}$ of etching is quite different from that observed in the low-nitrogen sector of the HPHT sample and in the CVD sample.  Considering just the HPHT sample, before etching, the total PL intensity is 44 times higher in the high-nitrogen sector than in the low-nitrogen sector, even through the implantation and annealing conditions were identical.  As the sample is etched, the PL intensity in the high-nitrogen sector rapidly decreases, indicating that most of the NV centers exist within the first $200\,\mathrm{nm}$, while the PL intensity in the low-nitrogen sector is approximately constant, suggesting that the first $200\,\mathrm{nm}$ of material is devoid of optically active NV centers.  This is a strong indication that, for the high-nitrogen sector, the NV concentration produced through implantation and annealing was limited mainly by available vacancies, while for the low-nitrogen sector the NV concentration was limited by the available nitrogen.  However, at an etch depth of $400\,\mathrm{nm}$, the PL intensities in the two sectors are almost equal (they differ by factor of 1.7), and at this point the NV concentration is probably limited by the available vacancies in both sectors.  As the etching progresses further, the weak background NV signals become important.  Since the high-nitrogen sector has a higher background NV concentration, the PL intensities in the two sectors again diverge for the largest etch depths.

\section{NV Concentration Estimate}

The PL measurement integrates contributions from all of the remaining NV centers produced by ion implantation and annealing, since the depth of focus of our optical setup is much larger than $1 \, \mu\mathrm{m}$ in diamond, as discussed above.  To a first approximation, the NV concentration is therefore proportional to the first derivative of the PL intensity with respect to etch depth $\zeta$.  However, there are two effects that complicate the analysis: spontaneous emission modification by the nearby diamond-air interface, and electronic depletion.  Spontaneous emission modification by a dielectric interface is well understood and can be included in the data analysis if we make an assumption about the radiative quantum efficiency of the NV center.  In this section, we describe how to include spontaneous emission modification and obtain density estimates that are corrected for this effect.  Electronic depletion is more complicated, but appears to be important only for the low-nitrogen material.  Electronic depletion is discussed and modeled in the following sections.

The spontaneous emission rate of a quantum dipole is proportional to the total power radiated by an equivalent classical dipole.  In the presence of a dielectric interface, the radiated power includes an interference term due to radiation reflecting off of the interface and performing work on the dipole.  We calculated the ratio between the total radiated power, including the dielectric interface, and that in bulk diamond, using a formula derived in Ref.~\onlinecite{ref:lukosz1977leb}.  This ratio is equal to the spontaneous emission rate enhancement factor $F_p(z)$ ($z$ is the depth below the surface) plotted in Fig~\ref{fig:NVdensity}.  For the NV center, we must include two dipoles that are orthogonal to the NV axis and to each other.  The NV axis can be oriented along any of the four $\langle 111 \rangle$ crystal directions, and in our samples the surface is $(100)$.  For this geometry, the radiated power averaged over the two dipoles (regardless of how they are chosen) gives the same result as an isotropic mixture of dipole moments.  For the result in Fig.~\ref{fig:NVdensity}(a), we have also averaged over $650-750\,\mathrm{nm}$ emission wavelengths, to take into account the broad NV$^-$ spectrum that is dominated by phonon sidebands.  The spontaneous emission enhancement (or suppression if $F_p<1$) is due entirely to the reflected wave, and only the radiated power into the substrate is modified.  The radiated power into air is unchanged, but as $F_p$ increases, the spontaneous emission lifetime of the NV center decreases, and thus the probability per excitation event for a photon to be emitted into air decreases.  In the limit of weak excitation at $532\,\mathrm{nm}$, the rate of excitation events does not depend on this spontaneous emission modification effect, and depends only on the driving electric field intensity, the dipole moment of the optical transition, and the spectral width of the phonon sidebands.  Thus, in the weak excitation limit, assuming a radiative efficiency of unity, the collection efficiency is inversely proportional to $F_p$.  We have defined a theoretical, relative collection efficiency $\eta=1/F_p$, plotted in Fig.~\ref{fig:NVdensity}(a).  We have ignored one subtlety by using a single $\eta$ based on the average radiated power from the two mutually orthogonal dipoles of the NV center.  If there were no relaxation between excited states, then the two dipoles would each have a different spontaneous emission lifetime, and we would have to use $\eta = (F_{p,1}^{-1} + F_{p,2}^{-1})/2$.  However, at room temperature, where rapid thermalization between the two orbital excited states may be expected, we believe it is more appropriate to average $F_p$ before taking the inverse.

\begin{figure}[ht]
\centering
\includegraphics[width=3.5in]{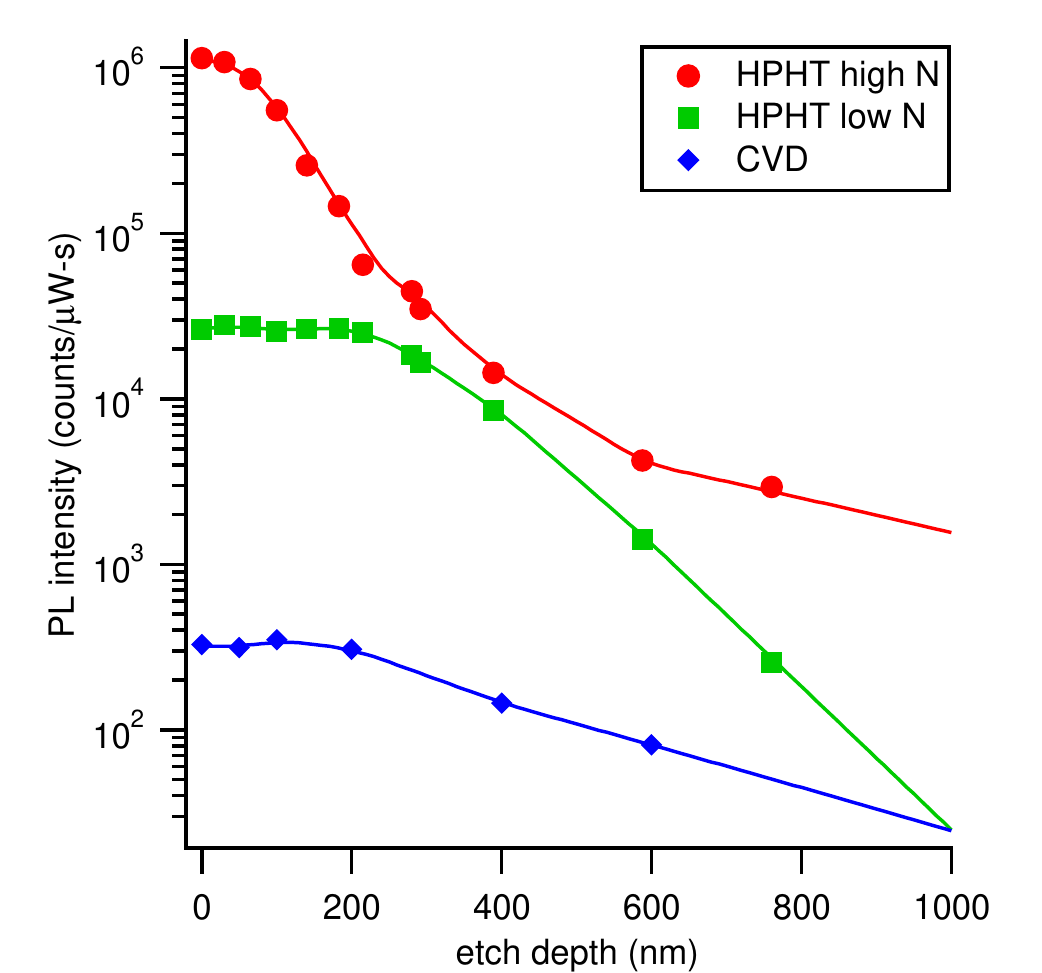}
\caption{Integrated PL intensity vs. etch depth for the HPHT sample, high-nitrogen sector (red circles) and low-nitrogen sector (green squares), and the CVD sample (blue diamonds).  For the CVD sample, the plotted intensity is the average intensity measured within the implanted squares minus the average intensity outside of the implanted squares (see text). Curves: PL intensity computed from the NV concentration estimates shown in Fig.~\ref{fig:NVdensity}.}
\label{fig:APDplots}
\end{figure}

\begin{figure}[ht]
\centering
\includegraphics[width=3.0in]{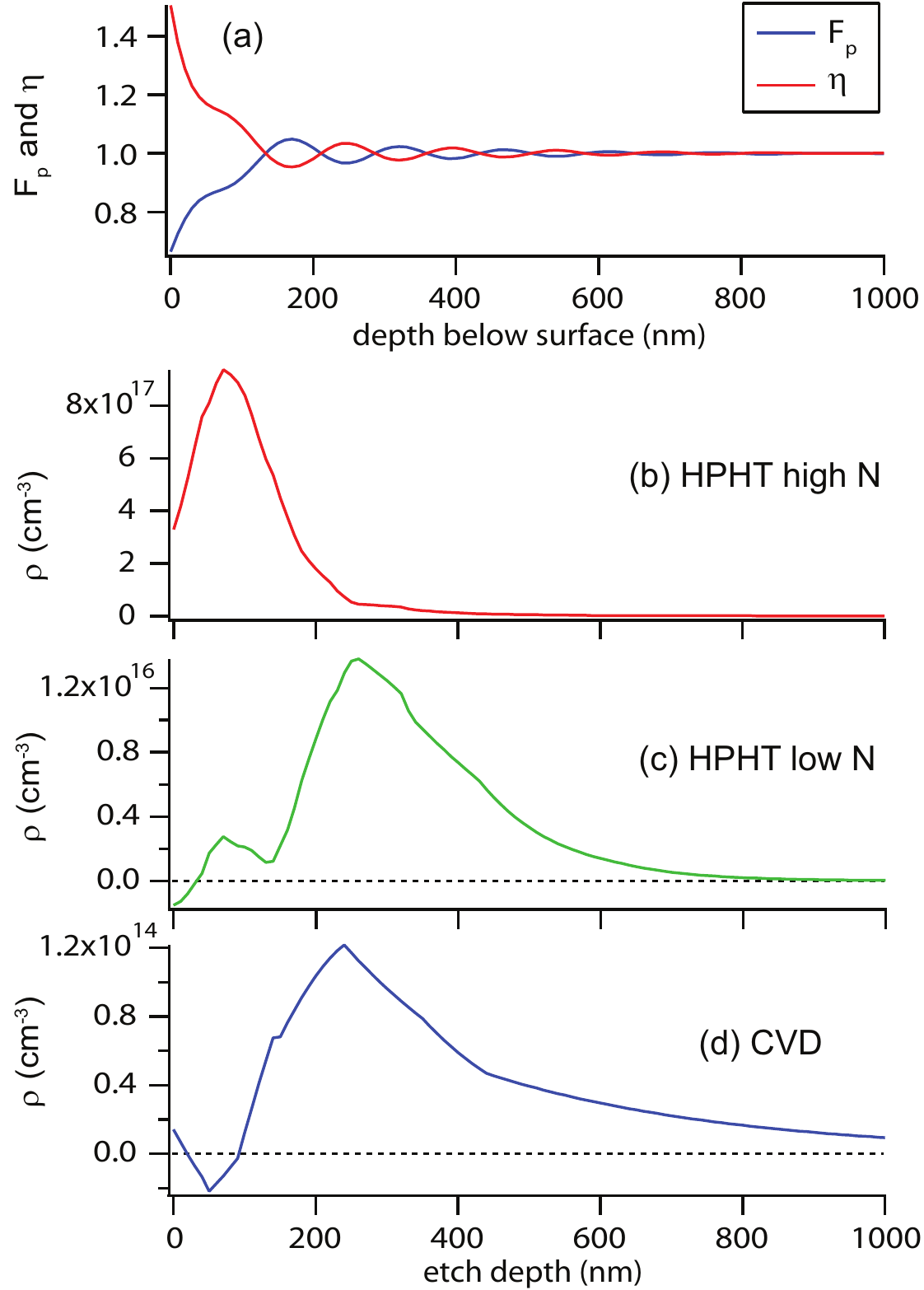}
\caption{(a) Calculated Purcell factor (blue) and corresponding relative collection efficiency (red) for a $[111]$-oriented NV center close to a $(100)$ diamond surface. (b-d) NV concentration vs. depth, estimated from the data in Fig.~\ref{fig:APDplots}.}
\label{fig:NVdensity}
\end{figure}

 Taking spontaneous emission modification into account, we expect the measured intensity $\sigma(\zeta)$, where $\zeta$ is the etch depth, to depend on $\eta(z)$ and the initial NV concentration $\rho(z)$ (before etching) according to,
\begin{equation}
\sigma(\zeta \ge 0) = \int_{\zeta}^\infty dz \, \eta(z-\zeta) \rho(z) \, .
\label{eq:sigma}
\end{equation}
Thus, to obtain $\rho(z)$ we must solve a deconvolution problem.  For our particular $\eta(z)$ a converging iterative solution can be obtained as follows:
\begin{eqnarray}
\rho(z) &=& \sum_{n=0}^\infty \rho_n(z) \, , \label{eq:rho1}  \\
\rho_0(z) &=& -\frac{1}{\eta(0)} \left. \frac{d\sigma}{d\zeta} \right|_{\zeta=z} \, , \label{eq:rho2} \\
\rho_{n>0}(z) &=& -\frac{1}{\eta(0)} \int_{z}^\infty d z_2 \eta^{\prime}(z_2-z) \rho_{n-1}(z_2) \, , \label{eq:rho3} \,
\end{eqnarray}
where $\eta^{\prime}(z) = d\eta/dz$.  The observed rate of convergence was such that, after 10 iterations, the remaining corrections had $\mathrm{Max} |\rho_n(z)|/ \mathrm{Max} |\rho_0(z)| < 10^{-5}$.  In analyzing the data we must also contend with a large amount of noise.  We believe this noise is due mainly to errors in replacing the sample before each measurement and in setting the focus, and we also have uncertainty in estimating the etch depth.  The observed fluctuations are much larger than the Poisson detection limit and are approximately 10\% of the count rate.  We used the following procedure to estimate $\rho(z)$: (1) perform a linear interpolation of $\log \sigma(z)$, and exponentiate to obtain a continuous version of $\rho(z)$.  (2) Calculate the corresponding $\rho(z)$ using Eqs.~\ref{eq:rho1}-\ref{eq:rho3}.  (3) Smooth the result using a moving average with $100 \,\mathrm{nm}$ total width to obtain $\bar{\rho}(z)$. (4) Calculate a $\bar{\sigma}(\zeta)$ from $\bar{\rho}(z)$ using Eq.~\ref{eq:sigma}, and compare with the original data.  The calculated $\bar{\rho}(z)$ are plotted in Fig.~\ref{fig:NVdensity}, and the corresponding $\bar{\sigma}(\zeta)$ are the fits shown in Fig.~\ref{fig:APDplots}.  Including spontaneous emission modification had a noticeable effect on the calculated NV profiles, but the results are qualitatively similar to those obtained simply by differentiating $\sigma(\zeta)$.  In order to estimate the absolute NV concentrations shown in the figure, we also made use of separate calibration measurements performed on single NV centers, where the count rate per unit excitation power was approximately $1.5 \times 10^4 \,\mathrm{s}^{-1} \mathrm{mW}^{-1}$, and the effective collection area of the confocal setup was determined to be approximately $0.5 \, \mu\mathrm{m}^2$.  The estimated sheet densities of optically active NV centers for the three measurements are $1.3 \times 10^{13} \, \mathrm{cm}^{-2}$ and $3.5 \times 10^{11} \, \mathrm{cm}^{-2}$ for the high- and low-nitrogen sectors, respectively, of the HPHT sample, and $4.3 \times 10^{9} \, \mathrm{cm}^{-2}$ for the CVD sample.

The results in Fig.~\ref{fig:NVdensity} confirm that, for the high-nitrogen sector of the HPHT sample, most of the optically active NV centers occur within the first $200\,\mathrm{nm}$.  The distribution is significantly broader than the predicted vacancy profile before annealing shown in Fig.~\ref{fig:implant}.  Such a spreading of the distribution might be expected because of vacancy diffusion during annealing.  However, for the measurements on low-nitrogen material, the behavior is quite different.  In these cases, the concentration of optically active NV centers is very low in the first $100\,\mathrm{nm}$, and the maximum concentration occurs approximately $250\,\mathrm{nm}$ below the surface.

\section{Spectroscopy Results}

\begin{figure}[ht]
\centering
\includegraphics[width=3in]{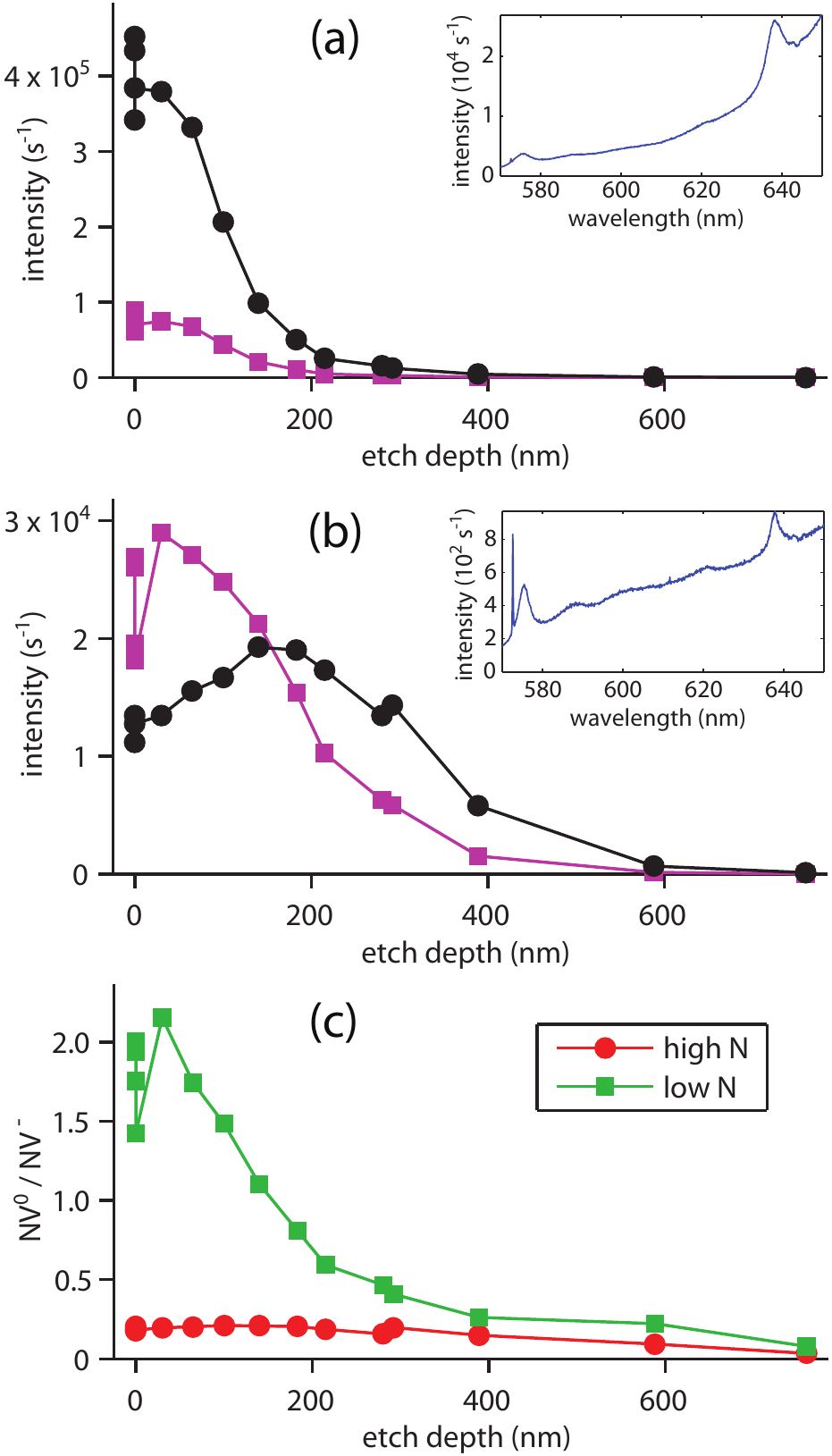}
\caption{(a,b)  Estimated NV$^0$ and NV$^-$ zero-phonon line (ZPL) intensities vs. etch depth for the (a) high-nitrogen and (b) low-nitrogen sectors of the HPHT sample.  Insets: example spectra showing the diamond Raman line at $572\,\mathrm{nm}$ and NV$^0$ and NV$^-$ zero-phonon lines (ZPLs) at $575\,\mathrm{nm}$ and $638\,\mathrm{nm}$, respectively. (c) The NV$^0$/NV$^-$ intensity ratio vs. etch depth for the high- and low-nitrogen sectors of the Sumitomo sample.}
\label{fig:spectra}
\end{figure}

Next, we present spectroscopy results from the HPHT sample that help explain the lack of PL originating near the surface in the low-nitrogen material.  The insets of Fig.~\ref{fig:spectra}(a,b) show typical room-temperature PL spectra from the high- and low-nitrogen sectors.  The most important features are the $575\,\mathrm{nm}$ and $638\,\mathrm{nm}$ peaks which have been identified as the zero-phonon lines (ZPLs) of NV$^0$ and NV$^-$, respectively~\cite{ref:mita1996cas}.  The sharp line at $572\,\mathrm{nm}$ that appears prominently in Fig.~\ref{fig:spectra}(b) is the diamond Raman line.  To extract the NV$^-$ and NV$^0$ PL intensities from each spectrum, we fitted a Voigt approximation~\cite{ref:liu2001sea} to the $575\,\mathrm{nm}$ and $638\,\mathrm{nm}$ peaks after subtracting a linearly sloped background fitted to regions on the two sides of the peak.  The extracted peak areas are plotted as a function of etch depth in the main plots of Figs.~\ref{fig:spectra}(a,b).  The NV$^0$/NV$^-$ intensity ratio for both sectors is plotted in Fig.~\ref{fig:spectra}(c).

For the first few hundred nanometers of etching, quite different behavior is observed for the high-nitrogen and low-nitrogen sectors.  In the high-nitrogen sector, the NV$^0$/NV$^-$ ratio is approximately 0.2 and changes little as the sample is etched, although it eventually decreases.  However, for the low-nitrogen sector, this ratio changes dramatically as the sample is etched, starting out at 2.2 after the first successful etch, and decreasing to 0.3 after $400\,\mathrm{nm}$ of etching.  Even more surprisingly, for the first $150\,\mathrm{nm}$ of etching the NV$^-$ intensity actually increases. The measured PL intensity includes the contributions from NV centers at all depths, and we do not expect significant absorption in the top layer.  We believe the most likely explanation is that the top layer interacts electronically with the layers below.  A simple model describing this interaction is presented in the next section.

Finally, we briefly address the low-temperature spectral linewidth in the implanted samples, a critical issue for some potential applications.  In the ideal case, the linewidth of NV centers created through ion implantation and annealing would be equal to the linewidth observed in background NV centers distributed throughout the sample.  For another Element 6 CVD sample with nitrogen content similar to that of the sample used in the etching experiment, the low-temperature linewidth (FWHM) of the NV$^-$ ZPL at $637\,\mathrm{nm}$ was measured to be $0.035 \, \mathrm{nm}$ for the background NV centers.  For the NV centers created by ion implanation ($10^{12} \, \mathrm{cm}^{-2}$, $200 \, \mathrm{keV}$ Ga$^+$) followed by etching to remove the top $300\,\mathrm{nm}$, the linewidth was $0.057 \, \mathrm{nm}$, and the tails of the ZPL peak were more prominent than for the background NV centers.  For the HPHT sample, the high-nitrogen sectors show broad linewidths ranging from $0.5 - 1.9 \, \mathrm{nm}$.  In one of the special low-nitrogen sectors, a strain-split NV$^-$ ZPL with $0.13\,\mathrm{nm}$ total linewidth was observed under normal excitation conditions.  However, when collecting PL through a GaP waveguide using evanescent coupling~\cite{ref:fu2008cnv} a much broader linewidth ($1.0\,\mathrm{nm}$) along with a large NV$^0$/NV$^-$ ratio was observed at the same location.  This indicates that NV centers close to the surface had severely degraded properties.  We have performed low-temperature spectroscopy on many samples implanted by a variety of methods and plan to report the results elsewhere~\cite{ref:aharonovich2008}.

\section{Simple Depletion Model}

In this section we seek to explain the behavior in Fig.~\ref{fig:spectra}(b), especially the initial increase in the NV$^-$ signal observed during the first $150\,\mathrm{nm}$ of etching. In our view the most likely explanation for this behavior is an electronic interaction, and in particular the presence of an electronic acceptor layer near the surface that removes electrons that would otherwise be available to form NV$^-$ in the layers underneath.

It has been shown that the relative PL intensities of NV$^0$ and NV$^-$ depend on the concentrations of impurities such as nitrogen, and on the optical excitation conditions~\cite{ref:mita1996cas, ref:iakoubovskii2000pvr, ref:wotherspoon2003pap, ref:manson2005pin}.  Several types of impurities are present that can act as electronic donors and acceptors.  Nitrogen impurities are expected to be the main electron donor, with an energy level $1.7\,\mathrm{eV}$ below the conduction band minimum~\cite{ref:farrer1969snd}.  The NV center itself is known to exist in the NV$^-$ and NV$^0$ states, and perhaps could also exist in an NV$^+$ state.  Because NV$^-$ has discrete excited states $1.94\,\mathrm{eV}$ above the ground states, the ground states of NV$^-$ must be $>1.94\,\mathrm{eV}$ below the conduction band minimum~\cite{ref:mita1996cas}. An abrupt change reported in the NV$^0$/NV$^-$ ratio at an excitation wavelength of $480 \, \mathrm{nm}$~\cite{ref:steeds2000pmt} suggests that the actual position of the NV$^-$ ground state is $2.58 \, \mathrm{eV}$ below the conduction band minimum.  Finally, we can only explain the results in Fig.~\ref{fig:spectra} if some species of electron acceptor is present in the top layer.  As discussed below, there are several possible acceptor species. For the present discussion we label the acceptor as A$^-$, and choose an energy level $1.4\,\mathrm{eV}$ above the valence band maximum as expected either for substitutional Ga impurities~\cite{ref:goss2005vic} or for graphitic defects~\cite{ref:ristein2000epd}.  The indirect bandgap of diamond is $E_g = E_c - E_v = 5.5\,\mathrm{eV}$.  Theoretically, for a junction between a strongly ``p-type'' layer and an ``n-type'' layer (nitrogen as the predominant impurity), a ``depletion layer'' of ionized nitrogen impurities is expected to have a width of approximately $x_n = \sqrt{2 \epsilon \Delta E / [N]e} = 90 \, \mathrm{nm}$ if the nitrogen concentration is $1 \, \mathrm{ppm}$.  Thus it is quite possible that a depletion layer could be responsible for a reduced NV$^-$ concentration before etching in the low-nitrogen material. For a nitrogen concentration of $100 \, \mathrm{ppm}$, the theoretical depletion width is only $9 \, \mathrm{nm}$, and interactions between different layers would be difficult to observe.

 As a first step toward quantitative modeling of the spectroscopy results, we have implemented a simple model that solves the 1-D Poisson equation,
\begin{equation}
\frac{d^2}{dz^2} \phi = -\frac{q}{\epsilon}
= \frac{e}{\epsilon} \left( n - [\mathrm{N}^+] + [\mathrm{NV}^-] + [\mathrm{A}^-] - p \right) \, ,
\label{eq:poisson}
\end{equation}
under the condition that the electronic occupations of each type of impurity, as well as the conduction-band electron concentration $n$ and valence-band hole concentration $p$, are in thermal equilibrium according to Fermi-Dirac statistics.  Eq.~\ref{eq:poisson} is then solved numerically under the constraint that the electric field is zero at $z=\pm \infty$, or equivalently, that the total charge is zero.  In the actual experiment, we do {\it not} necessarily expect the electronic populations to be in thermal equilibrium, since the sample is subject to optical excitation at $532\,\mathrm{nm}$.  Nevertheless, given our lack of knowledge about all of the optical excitation rates, this model is a good starting point.
In order to simulate the experiment, we must also assume some distribution of nitrogen impurities, NV centers and acceptors.  As an example, we have simulated the following distribution, where $z \ge 0$ is the depth below the original surface:
\begin{eqnarray}
\left[ \mathrm{N} \right] &=& \left[ \mathrm{N}^0 \right] + \left[ \mathrm{N}^+ \right] = 
  \mathrm{N}_\mathrm{bkg} - [\mathrm{NV}] \, , \\
\left[ \mathrm{NV} \right] &=& \left[ \mathrm{NV}^- \right] + \left[ \mathrm{NV}^0 \right] =
  \frac{r(z)}{1 + r(z)} \mathrm{N}_\mathrm{bkg}  \, , \\
r(z) &=& r_0 \left(1 + \theta(z-z_0) (e^{-(z-z_0)/\Delta z} - 1) \right) \, , \\
\left[ \mathrm{A} \right] &=& \left[ \mathrm{A}^- \right] + \left[ \mathrm{A}^0 \right] =
  a_0 \theta(z_1-z) \, ,
\end{eqnarray}
where $\mathrm{N}_\mathrm{bkg}$ is the initial nitrogen concentration (assumed to be uniform), $r(z)$ is a dimensionless rate for conversion of N to NV with a depth profile controlled by the parameters $r_0$, $z_0$ and $\Delta z$, $\theta(z)$ is the unit step function, and $a_0$ is the concentration of acceptors in a top layer of thickness $z_1$.  We have chosen this impurity distribution because it is simple and yet contains the most essential features of the NV distribution and acceptor layer as discussed above.

\begin{figure}[ht]
\centering
\includegraphics[width=3.3in]{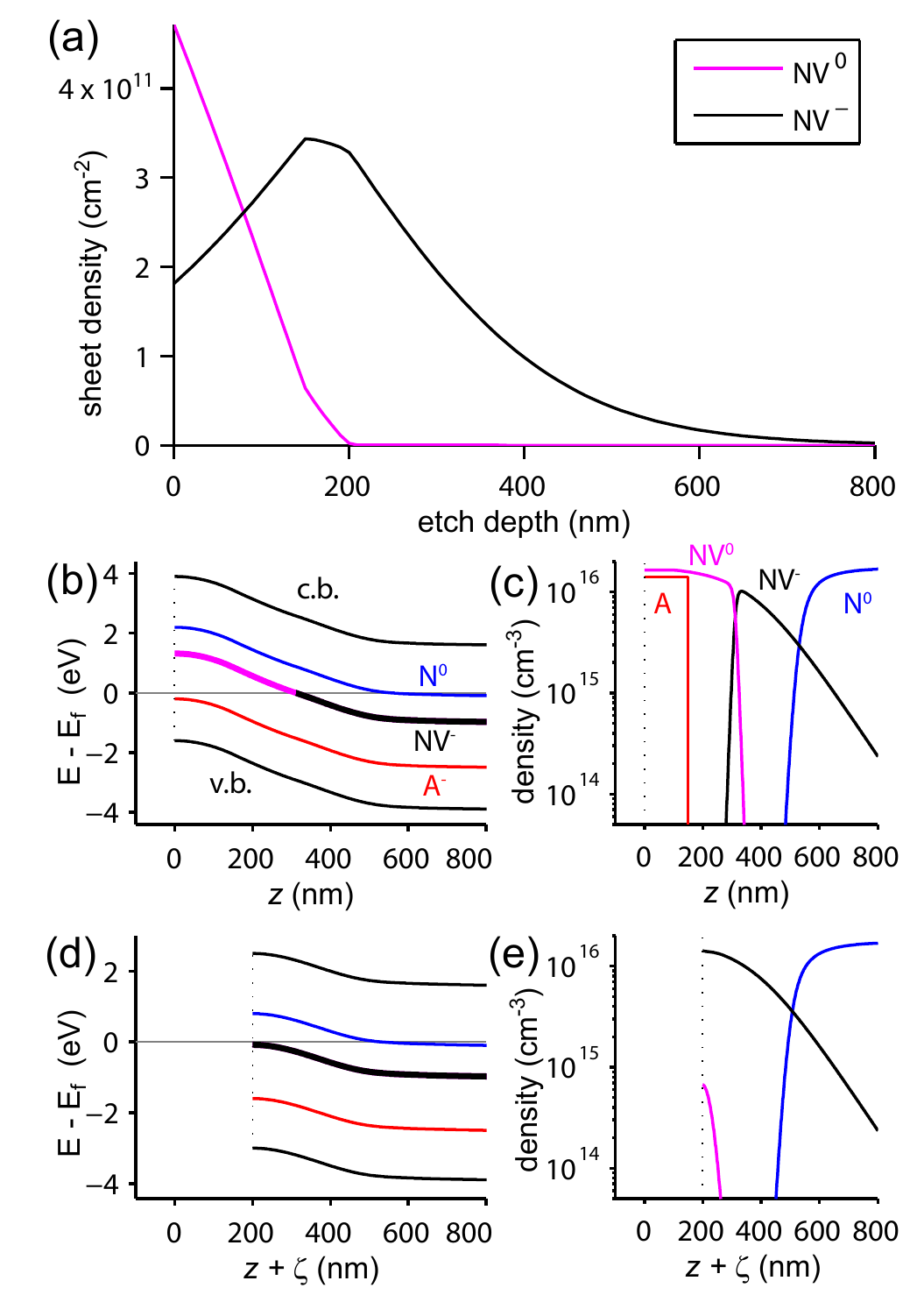}
\caption{Simulation of electron statistics using a thermal equilibrium model (see text for parameters):
(a) Total NV$^0$ and NV$^-$ populations (integrated over $z$) vs. etch depth.
(b) Band diagram before etching, showing the conduction band minimum (c.b.), valence band maximum (v.b.), nitrogen donor level (N$^0$), NV level (NV$^-$), and acceptor level (A$^-$), relative to the Fermi level (at zero energy).
(c) Populations before etching of neutral nitrogen impurities (N$^0$), negatively charged NV centers (NV$^-$), neutral NV centers (NV$^0$), and acceptors (A) vs. depth $z$.
(d) Band diagram after etching $200\,\mathrm{nm}$.
(e) Populations after etching $200\,\mathrm{nm}$.
}
\label{fig:simulations}
\end{figure}

The simulations shown in Fig.~\ref{fig:simulations} used the parameters $\mathrm{N}_\mathrm{bkg} = 1.76 \times 10^{16} \, \mathrm{cm}^{-3}$ ($0.1\,\mathrm{ppm}$), $r_0=15$, $z_0 = \Delta z = 100\,\mathrm{nm}$, $z_1 = 150\,\mathrm{nm}$, and $a_0 = 1.4 \times 10^{16} \, \mathrm{cm}^{-3}$.  The temperature was chosen to be $300\,\mathrm{K}$, which produces essentially the same results as $0\,\mathrm{K}$.  Fig.~\ref{fig:simulations}(a) shows the calculated 2-D concentrations of NV$^-$ and NV$^0$ (the 3-D concentration integrated over $z$) as a function of etch depth.  For these parameters, the simulation produces a similar behavior as observed in the spectroscopy data, where the NV$^-$ PL intensity increases at first, and then decreases.  To help in explaining this behavior, energy band diagrams and impurity electron concentrations are shown before etching (b,c) and after etching $200\,\mathrm{nm}$ (d,e).  Before etching, the $150\,\mathrm{nm}$ p-type layer takes electrons from nearby NV centers and nitrogen donors, so that NV centers cannot exist in their negatively charged state to a depth of $300 \,\mathrm{nm}$.  Below this depth, NV$^-$ can exist, but the total NV concentration there is low.  However, after removal of the p-type layer, nearly all of the remaining NV centers can exist in their negatively charged state.

Thus, through a suitable choice of parameters, we can obtain good qualitative agreement with the spectroscopy results shown in Fig.~\ref{fig:spectra}(b) for the low-nitrogen sector, demonstrating that electronic depletion is a plausible explanation for the initial rise in the NV$^-$ signal.  Such agreement is possible only for a limited range of parameters.  A substantial increase in NV$^-$ due to removal of the p-type layer is possible only if $\mathrm{N}_\mathrm{bkg}$ is small, and $\mathrm{N}_\mathrm{bkg}$ cannot be much larger than the value of $0.1\,\mathrm{ppm}$ used here.  This in turn limits the total number of NV centers that can exist in the model.  The maximum sheet density of NV$^-$ obtained in the model, $3.4 \times 10^{11} \, \mathrm{cm}^{-2}$, is in good agreement with the experiment: as discussed above, we estimated $3.5 \times 10^{11} \, \mathrm{cm}^{-2}$.  In order for the NV$^-$ population to increase gradually as the p-type layer is etched, the acceptor concentration must also be small, comparable to the initial nitrogen concentration in this case.  If it is much higher, the NV$^-$ population remains approximately constant until the p-type layer is almost completely removed, at which point the NV$^-$ population abruptly increases.  The acceptor concentration used here, approximately $0.8 \, \mathrm{ppm}$ in a $150\,\mathrm{nm}$ layer, corresponds to a sheet density of only $2.1 \times 10^{11} \, \mathrm{cm}^{-2}$, or only 7\% of the implanted Ga dose.

One important difference between theory and experiment can be seen in the NV$^0$/NV$^-$ ratio.  In the model, this ratio falls rapidly to zero once the damaged layer has been removed, and the NV concentration in the remaining layers is less than the nitrogen concentration.  In the experiment, this ratio reaches a minimum of approximately 0.25, and in general we observe this ratio to be dependent on the optical excitation power.  This is a limitation of the thermal equilibrium model presented above, especially at low temperature.  As the temperature in the model is increased (a crude way to simulate intense optical excitation), the ratios will change, and the equilibrium value will eventually depend on the effective masses of electrons and holes in the conduction and valence bands.  To model the system more accurately, we would need to use a rate-equation model for optical excitation combined with drift and diffusion of electrons and holes in the conduction and valance bands.  Such a non-equilibrium model is feasible to solve numerically, but the main difficulty is that we have little knowledge of the ionization rates of nitrogen impurities, NV centers, and acceptors under optical excitation, as well as the electron and hole capture rates for these impurities, all of which are important.  We have observed time-dependent behavior in the low-nitrogen sector that gives some hint of these rates: for excitation with $4.5\,\mu\mathrm{W}$ at $532\,\mathrm{nm}$ we observed that if the excitation is suddenly switched on, the PL intensity is initially smaller and increases by a factor of two with $\sim 70\,\mathrm{ms}$ rise-time.  This rise time could be associated, for example, with optically induced charge transfer from acceptors to NV centers in the depleted region by means of free carriers.

Finally, let us briefly address the question of what acceptor species could be responsible for creating such a depletion layer.  Several possibilities exist.  Damage close to the surface, created by polishing, for example, could have been present even before ion implantation.  If such damage consists of graphitic defects, this could pin the Fermi level $1.4\,\mathrm{eV}$ above the valence band maximum~\cite{ref:ristein2000epd}.  However, such defects would have to be distributed throughout the top $\sim 150\,\mathrm{nm}$ of diamond, in a low concentration, to explain the observed result.  Ion implantation introduces vacancies, self-interstitials, and Ga impurities.  The subsequent annealing process is complex, and the final end-products are uncertain.  Monovacancies could remain, but with an acceptor level apparently positioned $2.85\,\mathrm{eV}$ above the valence band maximum~\cite{ref:dannefaer2001pas} (nearly identical to NV$^-$), it is difficult to explain the observed results in terms of this acceptor.  However, if vacancies combine to form extended defects~\cite{ref:hounsome2006obc}, a high concentration of acceptor states throughout the diamond bandgap could exist.  Finally, it is quite possible that the implanted Ga atoms are the dominant acceptor.  Initially we expected that Ga impurities would not be important, since according to the simulation in Fig.~\ref{fig:implant}, $\sim 770$ vacancies are produced for each implanted Ga$^+$ ion.  However, the sheet density of acceptors needed to produce a depletion layer in our model is only 7\% of the implanted Ga$^+$ dose.  Acceptor levels for substitutional Ga impurities and for Ga-vacancy complexes have recently been calculated, with positions $1.4 \, \mathrm{eV}$ and $1.7 \, \mathrm{eV}$ above the valence band maximum, respectively~\cite{ref:goss2005vic}.

\section{Conclusions}

From the data and simulations presented above, we can draw several conclusions related to formation of NV centers through implantation of heavy ions followed by annealing.  First, the vertical distance over which vacancies diffuse to form NV centers is only a few hundred nanometers, even for a fairly high annealing temperature of $925^{\circ} \, \mathrm{C}$.  Second, if the implantation dose is chosen to maximize the PL intensity in diamond with a moderately low nitrogen concentration ($\sim 1 \, \mathrm{ppm}$), most of the NV$^-$ PL that is produced originates from NV centers $>200\,\mathrm{nm}$ below the surface, while NV centers closer to the surface are either in the NV$^0$ state or in other states that do not emit in our detection band.  Such a situation is not useful for coupling NV centers to optical structures.  Finally, the spectroscopic results in the low-nitrogen diamond are highly suggestive of electronic depletion effects.  Our model shows that the observed behavior can be explained by the presence of electronic acceptors near the surface, which could consist of impurities or defects introduced through the ion implantation and annealing, or damage originally present in the sample.  Controlling such effects will be important in finding improved methods to couple dense ensembles of NV centers to optical structures on a diamond surface, and also for fabrication of single negatively-charged NV centers in ultra-pure diamond, where electronic depletion layers thicker than $1\,\mu \mathrm{m}$ are possible.

The authors thank Dirk Englund and Steven Prawer for helpful discussions.


\begin{thebibliography}{53}
\expandafter\ifx\csname natexlab\endcsname\relax\def\natexlab#1{#1}\fi
\expandafter\ifx\csname bibnamefont\endcsname\relax
  \def\bibnamefont#1{#1}\fi
\expandafter\ifx\csname bibfnamefont\endcsname\relax
  \def\bibfnamefont#1{#1}\fi
\expandafter\ifx\csname citenamefont\endcsname\relax
  \def\citenamefont#1{#1}\fi
\expandafter\ifx\csname url\endcsname\relax
  \def\url#1{\texttt{#1}}\fi
\expandafter\ifx\csname urlprefix\endcsname\relax\def\urlprefix{URL }\fi
\providecommand{\bibinfo}[2]{#2}
\providecommand{\eprint}[2][]{\url{#2}}

\bibitem[{\citenamefont{van Oort et~al.}(1988)\citenamefont{van Oort, Manson,
  and Glasbeek}}]{ref:vanoort1988ods}
\bibinfo{author}{\bibfnamefont{E.}~\bibnamefont{van Oort}},
  \bibinfo{author}{\bibfnamefont{N.}~\bibnamefont{Manson}}, \bibnamefont{and}
  \bibinfo{author}{\bibfnamefont{M.}~\bibnamefont{Glasbeek}},
  \bibinfo{journal}{J. Phys. C} \textbf{\bibinfo{volume}{21}},
  \bibinfo{pages}{4385} (\bibinfo{year}{1988}).

\bibitem[{\citenamefont{Jelezko
  et~al.}(2004{\natexlab{a}})\citenamefont{Jelezko, Gaebel, Popa, Gruber, and
  Wrachtrup}}]{ref:jelezko2004obs}
\bibinfo{author}{\bibfnamefont{F.}~\bibnamefont{Jelezko}},
  \bibinfo{author}{\bibfnamefont{T.}~\bibnamefont{Gaebel}},
  \bibinfo{author}{\bibfnamefont{I.}~\bibnamefont{Popa}},
  \bibinfo{author}{\bibfnamefont{A.}~\bibnamefont{Gruber}}, \bibnamefont{and}
  \bibinfo{author}{\bibfnamefont{J.}~\bibnamefont{Wrachtrup}},
  \bibinfo{journal}{Phys. Rev. Lett.} \textbf{\bibinfo{volume}{92}},
  \bibinfo{pages}{076401} (\bibinfo{year}{2004}{\natexlab{a}}).

\bibitem[{\citenamefont{Hanson et~al.}(2006)\citenamefont{Hanson, Gywat, and
  Awschalom}}]{ref:hanson2006rtm}
\bibinfo{author}{\bibfnamefont{R.}~\bibnamefont{Hanson}},
  \bibinfo{author}{\bibfnamefont{O.}~\bibnamefont{Gywat}}, \bibnamefont{and}
  \bibinfo{author}{\bibfnamefont{D.}~\bibnamefont{Awschalom}},
  \bibinfo{journal}{Phys. Rev. B} \textbf{\bibinfo{volume}{74}},
  \bibinfo{pages}{161203} (\bibinfo{year}{2006}).

\bibitem[{\citenamefont{Gaebel et~al.}(2006)\citenamefont{Gaebel, Domhan, Popa,
  Wittmann, Neumann, Jelezko, Rabeau, Stravrias, Greentree, Prawer, Meijer,
  Twamley, Hemmer, and Wrachtrup}}]{ref:gaebel2006rtc}
\bibinfo{author}{\bibfnamefont{T.}~\bibnamefont{Gaebel}},
  \bibinfo{author}{\bibfnamefont{M.}~\bibnamefont{Domhan}},
  \bibinfo{author}{\bibfnamefont{I.}~\bibnamefont{Popa}},
  \bibinfo{author}{\bibfnamefont{C.}~\bibnamefont{Wittmann}},
  \bibinfo{author}{\bibfnamefont{P.}~\bibnamefont{Neumann}},
  \bibinfo{author}{\bibfnamefont{F.}~\bibnamefont{Jelezko}},
  \bibinfo{author}{\bibfnamefont{J.~R.} \bibnamefont{Rabeau}},
  \bibinfo{author}{\bibfnamefont{N.}~\bibnamefont{Stravrias}},
  \bibinfo{author}{\bibfnamefont{A.~D.} \bibnamefont{Greentree}},
  \bibinfo{author}{\bibfnamefont{S.}~\bibnamefont{Prawer}},
  \bibinfo{author}{\bibfnamefont{J.}~\bibnamefont{Meijer}},
  \bibinfo{author}{\bibfnamefont{J.}~\bibnamefont{Twamley}},
  \bibinfo{author}{\bibfnamefont{P.~R.} \bibnamefont{Hemmer}},
  \bibnamefont{and}
  \bibinfo{author}{\bibfnamefont{J.}~\bibnamefont{Wrachtrup}},
  \bibinfo{journal}{Nat. Phys.} \textbf{\bibinfo{volume}{2}},
  \bibinfo{pages}{408} (\bibinfo{year}{2006}).

\bibitem[{\citenamefont{Jelezko
  et~al.}(2004{\natexlab{b}})\citenamefont{Jelezko, Gaebel, Popa, Domhan,
  Gruber, and Wrachtrup}}]{ref:jelezko2004ocn}
\bibinfo{author}{\bibfnamefont{F.}~\bibnamefont{Jelezko}},
  \bibinfo{author}{\bibfnamefont{T.}~\bibnamefont{Gaebel}},
  \bibinfo{author}{\bibfnamefont{I.}~\bibnamefont{Popa}},
  \bibinfo{author}{\bibfnamefont{M.}~\bibnamefont{Domhan}},
  \bibinfo{author}{\bibfnamefont{A.}~\bibnamefont{Gruber}}, \bibnamefont{and}
  \bibinfo{author}{\bibfnamefont{J.}~\bibnamefont{Wrachtrup}},
  \bibinfo{journal}{Phys. Rev. Lett.} \textbf{\bibinfo{volume}{93}},
  \bibinfo{pages}{130501} (\bibinfo{year}{2004}{\natexlab{b}}).

\bibitem[{\citenamefont{Childress et~al.}(2006)\citenamefont{Childress,
  Gurudev~Dutt, Taylor, Zibrov, Jelezko, Wrachtrup, Hemmer, and
  Lukin}}]{ref:childress2006cdc}
\bibinfo{author}{\bibfnamefont{L.}~\bibnamefont{Childress}},
  \bibinfo{author}{\bibfnamefont{M.~V.} \bibnamefont{Gurudev~Dutt}},
  \bibinfo{author}{\bibfnamefont{J.~M.} \bibnamefont{Taylor}},
  \bibinfo{author}{\bibfnamefont{A.~S.} \bibnamefont{Zibrov}},
  \bibinfo{author}{\bibfnamefont{F.}~\bibnamefont{Jelezko}},
  \bibinfo{author}{\bibfnamefont{J.}~\bibnamefont{Wrachtrup}},
  \bibinfo{author}{\bibfnamefont{P.~R.} \bibnamefont{Hemmer}},
  \bibnamefont{and} \bibinfo{author}{\bibfnamefont{M.~D.} \bibnamefont{Lukin}},
  \bibinfo{journal}{Science} \textbf{\bibinfo{volume}{314}},
  \bibinfo{pages}{281} (\bibinfo{year}{2006}).

\bibitem[{\citenamefont{{Gurudev Dutt} et~al.}(2007)\citenamefont{{Gurudev
  Dutt}, Childress, Jiang, Togan, Maze, Jelezko, Zibrov, Hemmer, and
  Lukin}}]{ref:gurudevdutt2007qrb}
\bibinfo{author}{\bibfnamefont{M.~V.} \bibnamefont{{Gurudev Dutt}}},
  \bibinfo{author}{\bibfnamefont{L.}~\bibnamefont{Childress}},
  \bibinfo{author}{\bibfnamefont{L.}~\bibnamefont{Jiang}},
  \bibinfo{author}{\bibfnamefont{E.}~\bibnamefont{Togan}},
  \bibinfo{author}{\bibfnamefont{J.}~\bibnamefont{Maze}},
  \bibinfo{author}{\bibfnamefont{F.}~\bibnamefont{Jelezko}},
  \bibinfo{author}{\bibfnamefont{A.~S.} \bibnamefont{Zibrov}},
  \bibinfo{author}{\bibfnamefont{P.~R.} \bibnamefont{Hemmer}},
  \bibnamefont{and} \bibinfo{author}{\bibfnamefont{M.~D.} \bibnamefont{Lukin}},
  \bibinfo{journal}{Science} \textbf{\bibinfo{volume}{316}},
  \bibinfo{pages}{1312} (\bibinfo{year}{2007}).

\bibitem[{\citenamefont{Kennedy et~al.}(2003)\citenamefont{Kennedy, Colton,
  Butler, Linares, and Doering}}]{ref:kennedy2003lct}
\bibinfo{author}{\bibfnamefont{T.~A.} \bibnamefont{Kennedy}},
  \bibinfo{author}{\bibfnamefont{J.~S.} \bibnamefont{Colton}},
  \bibinfo{author}{\bibfnamefont{J.~E.} \bibnamefont{Butler}},
  \bibinfo{author}{\bibfnamefont{R.~C.} \bibnamefont{Linares}},
  \bibnamefont{and} \bibinfo{author}{\bibfnamefont{P.~J.}
  \bibnamefont{Doering}}, \bibinfo{journal}{Appl. Phys. Lett.}
  \textbf{\bibinfo{volume}{83}}, \bibinfo{pages}{4190} (\bibinfo{year}{2003}).

\bibitem[{\citenamefont{Mizuochi et~al.}(2008)\citenamefont{Mizuochi, Neumann,
  Rempp, Beck, Jacques, Siyushev, Nakamura, Twitchen, Watanabe, Yamasaki,
  Jelezko, and Wrachtrup}}]{ref:mizuochi2008csp}
\bibinfo{author}{\bibfnamefont{N.}~\bibnamefont{Mizuochi}},
  \bibinfo{author}{\bibfnamefont{P.}~\bibnamefont{Neumann}},
  \bibinfo{author}{\bibfnamefont{F.}~\bibnamefont{Rempp}},
  \bibinfo{author}{\bibfnamefont{J.}~\bibnamefont{Beck}},
  \bibinfo{author}{\bibfnamefont{V.}~\bibnamefont{Jacques}},
  \bibinfo{author}{\bibfnamefont{P.}~\bibnamefont{Siyushev}},
  \bibinfo{author}{\bibfnamefont{K.}~\bibnamefont{Nakamura}},
  \bibinfo{author}{\bibfnamefont{D.}~\bibnamefont{Twitchen}},
  \bibinfo{author}{\bibfnamefont{H.}~\bibnamefont{Watanabe}},
  \bibinfo{author}{\bibfnamefont{S.}~\bibnamefont{Yamasaki}},
  \bibinfo{author}{\bibfnamefont{F.}~\bibnamefont{Jelezko}}, \bibnamefont{and}
  \bibinfo{author}{\bibfnamefont{J.}~\bibnamefont{Wrachtrup}},
  \bibinfo{journal}{eprint arXiv: 0811.4731}  (\bibinfo{year}{2008}).

\bibitem[{\citenamefont{Childress et~al.}(2005)\citenamefont{Childress, Taylor,
  Sorensen, and Lukin}}]{ref:childress2005ftq}
\bibinfo{author}{\bibfnamefont{L.}~\bibnamefont{Childress}},
  \bibinfo{author}{\bibfnamefont{J.~M.} \bibnamefont{Taylor}},
  \bibinfo{author}{\bibfnamefont{A.~S.} \bibnamefont{Sorensen}},
  \bibnamefont{and} \bibinfo{author}{\bibfnamefont{M.~D.} \bibnamefont{Lukin}},
  \bibinfo{journal}{Phys. Rev. A} \textbf{\bibinfo{volume}{72}},
  \bibinfo{pages}{52330} (\bibinfo{year}{2005}).

\bibitem[{\citenamefont{Nizovtsev et~al.}(2005)\citenamefont{Nizovtsev, Kilin,
  Jelezko, Gaebal, Popa, Gruber, and Wrachtrup}}]{ref:nizovtsev2005qcb}
\bibinfo{author}{\bibfnamefont{A.}~\bibnamefont{Nizovtsev}},
  \bibinfo{author}{\bibfnamefont{S.}~\bibnamefont{Kilin}},
  \bibinfo{author}{\bibfnamefont{F.}~\bibnamefont{Jelezko}},
  \bibinfo{author}{\bibfnamefont{T.}~\bibnamefont{Gaebal}},
  \bibinfo{author}{\bibfnamefont{I.}~\bibnamefont{Popa}},
  \bibinfo{author}{\bibfnamefont{A.}~\bibnamefont{Gruber}}, \bibnamefont{and}
  \bibinfo{author}{\bibfnamefont{J.}~\bibnamefont{Wrachtrup}},
  \bibinfo{journal}{Optics and Spectroscopy} \textbf{\bibinfo{volume}{99}},
  \bibinfo{pages}{233} (\bibinfo{year}{2005}).

\bibitem[{\citenamefont{Benjamin et~al.}(2006)\citenamefont{Benjamin, Browne,
  Fitzsimons, and Morton}}]{ref:benjamin2006bgs}
\bibinfo{author}{\bibfnamefont{S.~C.} \bibnamefont{Benjamin}},
  \bibinfo{author}{\bibfnamefont{D.~E.} \bibnamefont{Browne}},
  \bibinfo{author}{\bibfnamefont{J.}~\bibnamefont{Fitzsimons}},
  \bibnamefont{and} \bibinfo{author}{\bibfnamefont{J.~J.~L.}
  \bibnamefont{Morton}}, \bibinfo{journal}{New J. Phys.}
  \textbf{\bibinfo{volume}{8}}, \bibinfo{pages}{141} (\bibinfo{year}{2006}).

\bibitem[{\citenamefont{Taylor et~al.}(2008)\citenamefont{Taylor, Cappellaro,
  Childress, Jiang, Budker, Hemmer, Yacoby, Walsworth, and
  Lukin}}]{ref:taylor2008hsd}
\bibinfo{author}{\bibfnamefont{J.~M.} \bibnamefont{Taylor}},
  \bibinfo{author}{\bibfnamefont{P.}~\bibnamefont{Cappellaro}},
  \bibinfo{author}{\bibfnamefont{L.}~\bibnamefont{Childress}},
  \bibinfo{author}{\bibfnamefont{L.}~\bibnamefont{Jiang}},
  \bibinfo{author}{\bibfnamefont{D.}~\bibnamefont{Budker}},
  \bibinfo{author}{\bibfnamefont{P.~R.} \bibnamefont{Hemmer}},
  \bibinfo{author}{\bibfnamefont{A.}~\bibnamefont{Yacoby}},
  \bibinfo{author}{\bibfnamefont{R.}~\bibnamefont{Walsworth}},
  \bibnamefont{and} \bibinfo{author}{\bibfnamefont{M.~D.} \bibnamefont{Lukin}},
  \bibinfo{journal}{Nat. Phys.} \textbf{\bibinfo{volume}{4}},
  \bibinfo{pages}{810} (\bibinfo{year}{2008}).

\bibitem[{\citenamefont{Degen}(2008)}]{ref:degen2008smf}
\bibinfo{author}{\bibfnamefont{C.}~\bibnamefont{Degen}},
  \bibinfo{journal}{Appl. Phys. Lett.} \textbf{\bibinfo{volume}{92}},
  \bibinfo{pages}{243111} (\bibinfo{year}{2008}).

\bibitem[{\citenamefont{Balasubramanian
  et~al.}(2008)\citenamefont{Balasubramanian, Chan, Kolesov, Al-Hmoud, Tisler,
  Shin, Kim, Wojcik, Hemmer, Krueger, Hanke, Leitenstorfer, Bratschitsch,
  Jelezko, and Wrachtrup}}]{ref:balasubramanian2008nim}
\bibinfo{author}{\bibfnamefont{G.}~\bibnamefont{Balasubramanian}},
  \bibinfo{author}{\bibfnamefont{I.~Y.} \bibnamefont{Chan}},
  \bibinfo{author}{\bibfnamefont{R.}~\bibnamefont{Kolesov}},
  \bibinfo{author}{\bibfnamefont{M.}~\bibnamefont{Al-Hmoud}},
  \bibinfo{author}{\bibfnamefont{J.}~\bibnamefont{Tisler}},
  \bibinfo{author}{\bibfnamefont{C.}~\bibnamefont{Shin}},
  \bibinfo{author}{\bibfnamefont{C.}~\bibnamefont{Kim}},
  \bibinfo{author}{\bibfnamefont{A.}~\bibnamefont{Wojcik}},
  \bibinfo{author}{\bibfnamefont{P.~R.} \bibnamefont{Hemmer}},
  \bibinfo{author}{\bibfnamefont{A.}~\bibnamefont{Krueger}},
  \bibinfo{author}{\bibfnamefont{T.}~\bibnamefont{Hanke}},
  \bibinfo{author}{\bibfnamefont{A.}~\bibnamefont{Leitenstorfer}},
  \bibinfo{author}{\bibfnamefont{R.}~\bibnamefont{Bratschitsch}},
  \bibinfo{author}{\bibfnamefont{F.}~\bibnamefont{Jelezko}}, \bibnamefont{and}
  \bibinfo{author}{\bibfnamefont{J.}~\bibnamefont{Wrachtrup}},
  \bibinfo{journal}{Nature (London)} \textbf{\bibinfo{volume}{455}},
  \bibinfo{pages}{648} (\bibinfo{year}{2008}).

\bibitem[{\citenamefont{Maze et~al.}(2008)\citenamefont{Maze, Stanwix, Hodges,
  Hong, Taylor, Cappellaro, Jiang, Dutt, Togan, Zibrov, Yacoby, Walsworth, and
  Lukin}}]{ref:maze2008nms}
\bibinfo{author}{\bibfnamefont{J.~R.} \bibnamefont{Maze}},
  \bibinfo{author}{\bibfnamefont{P.~L.} \bibnamefont{Stanwix}},
  \bibinfo{author}{\bibfnamefont{J.~S.} \bibnamefont{Hodges}},
  \bibinfo{author}{\bibfnamefont{S.}~\bibnamefont{Hong}},
  \bibinfo{author}{\bibfnamefont{J.~M.} \bibnamefont{Taylor}},
  \bibinfo{author}{\bibfnamefont{P.}~\bibnamefont{Cappellaro}},
  \bibinfo{author}{\bibfnamefont{L.}~\bibnamefont{Jiang}},
  \bibinfo{author}{\bibfnamefont{M.~V.~G.} \bibnamefont{Dutt}},
  \bibinfo{author}{\bibfnamefont{E.}~\bibnamefont{Togan}},
  \bibinfo{author}{\bibfnamefont{A.~S.} \bibnamefont{Zibrov}},
  \bibinfo{author}{\bibfnamefont{A.}~\bibnamefont{Yacoby}},
  \bibinfo{author}{\bibfnamefont{R.~L.} \bibnamefont{Walsworth}},
  \bibnamefont{and} \bibinfo{author}{\bibfnamefont{M.~D.} \bibnamefont{Lukin}},
  \bibinfo{journal}{Nature (London)} \textbf{\bibinfo{volume}{455}},
  \bibinfo{pages}{644} (\bibinfo{year}{2008}).

\bibitem[{\citenamefont{Hiscocks et~al.}(2008)\citenamefont{Hiscocks, Ganesan,
  Gibson, Huntington, Ladouceur, and Prawer}}]{ref:hiscocks2008dwf}
\bibinfo{author}{\bibfnamefont{M.}~\bibnamefont{Hiscocks}},
  \bibinfo{author}{\bibfnamefont{K.}~\bibnamefont{Ganesan}},
  \bibinfo{author}{\bibfnamefont{B.}~\bibnamefont{Gibson}},
  \bibinfo{author}{\bibfnamefont{S.}~\bibnamefont{Huntington}},
  \bibinfo{author}{\bibfnamefont{F.}~\bibnamefont{Ladouceur}},
  \bibnamefont{and} \bibinfo{author}{\bibfnamefont{S.}~\bibnamefont{Prawer}},
  \bibinfo{journal}{Optics Express} \textbf{\bibinfo{volume}{16}},
  \bibinfo{pages}{19512} (\bibinfo{year}{2008}).

\bibitem[{\citenamefont{Fu et~al.}(2008)\citenamefont{Fu, Santori, Barclay,
  Aharonovich, Prawer, Meyer, Holm, and Beausoleil}}]{ref:fu2008cnv}
\bibinfo{author}{\bibfnamefont{K.}~\bibnamefont{Fu}},
  \bibinfo{author}{\bibfnamefont{C.}~\bibnamefont{Santori}},
  \bibinfo{author}{\bibfnamefont{P.}~\bibnamefont{Barclay}},
  \bibinfo{author}{\bibfnamefont{I.}~\bibnamefont{Aharonovich}},
  \bibinfo{author}{\bibfnamefont{S.}~\bibnamefont{Prawer}},
  \bibinfo{author}{\bibfnamefont{N.}~\bibnamefont{Meyer}},
  \bibinfo{author}{\bibfnamefont{A.}~\bibnamefont{Holm}}, \bibnamefont{and}
  \bibinfo{author}{\bibfnamefont{R.}~\bibnamefont{Beausoleil}},
  \bibinfo{journal}{Appl. Phys. Lett.} \textbf{\bibinfo{volume}{93}},
  \bibinfo{pages}{234107} (\bibinfo{year}{2008}).

\bibitem[{\citenamefont{Park et~al.}(2006)\citenamefont{Park, Cook, and
  Wang}}]{ref:park2006cqw}
\bibinfo{author}{\bibfnamefont{Y.-S.} \bibnamefont{Park}},
  \bibinfo{author}{\bibfnamefont{A.}~\bibnamefont{Cook}}, \bibnamefont{and}
  \bibinfo{author}{\bibfnamefont{H.}~\bibnamefont{Wang}},
  \bibinfo{journal}{Nano Letters} \textbf{\bibinfo{volume}{6}},
  \bibinfo{pages}{2075} (\bibinfo{year}{2006}).

\bibitem[{\citenamefont{Greentree et~al.}(2006)\citenamefont{Greentree,
  Olivero, Draganski, Trajkov, Rabeau, Reichart, Gibson, Rubanov, Huntington,
  Jamieson, and Prawer}}]{ref:greentree2006ccd}
\bibinfo{author}{\bibfnamefont{A.}~\bibnamefont{Greentree}},
  \bibinfo{author}{\bibfnamefont{P.}~\bibnamefont{Olivero}},
  \bibinfo{author}{\bibfnamefont{M.}~\bibnamefont{Draganski}},
  \bibinfo{author}{\bibfnamefont{E.}~\bibnamefont{Trajkov}},
  \bibinfo{author}{\bibfnamefont{J.}~\bibnamefont{Rabeau}},
  \bibinfo{author}{\bibfnamefont{P.}~\bibnamefont{Reichart}},
  \bibinfo{author}{\bibfnamefont{B.}~\bibnamefont{Gibson}},
  \bibinfo{author}{\bibfnamefont{S.}~\bibnamefont{Rubanov}},
  \bibinfo{author}{\bibfnamefont{S.}~\bibnamefont{Huntington}},
  \bibinfo{author}{\bibfnamefont{D.}~\bibnamefont{Jamieson}}, \bibnamefont{and}
  \bibinfo{author}{\bibfnamefont{S.}~\bibnamefont{Prawer}},
  \bibinfo{journal}{J. Phys. Cond. Mat.} \textbf{\bibinfo{volume}{18}},
  \bibinfo{pages}{825} (\bibinfo{year}{2006}).

\bibitem[{\citenamefont{Tomljenovic-Hanic
  et~al.}(2006)\citenamefont{Tomljenovic-Hanic, Steel, de~Sterke, and
  Salzman}}]{ref:tomljenovichanic2006dbp}
\bibinfo{author}{\bibfnamefont{S.}~\bibnamefont{Tomljenovic-Hanic}},
  \bibinfo{author}{\bibfnamefont{M.}~\bibnamefont{Steel}},
  \bibinfo{author}{\bibfnamefont{C.}~\bibnamefont{de~Sterke}},
  \bibnamefont{and} \bibinfo{author}{\bibfnamefont{J.}~\bibnamefont{Salzman}},
  \bibinfo{journal}{Optics Express} \textbf{\bibinfo{volume}{14}},
  \bibinfo{pages}{3556} (\bibinfo{year}{2006}).

\bibitem[{\citenamefont{Wang et~al.}(2007{\natexlab{a}})\citenamefont{Wang,
  Choi, Lee, Hu, Yang, and Butler}}]{ref:wang2007owg}
\bibinfo{author}{\bibfnamefont{C.~F.} \bibnamefont{Wang}},
  \bibinfo{author}{\bibfnamefont{Y.-S.} \bibnamefont{Choi}},
  \bibinfo{author}{\bibfnamefont{J.~C.} \bibnamefont{Lee}},
  \bibinfo{author}{\bibfnamefont{E.~L.} \bibnamefont{Hu}},
  \bibinfo{author}{\bibfnamefont{J.}~\bibnamefont{Yang}}, \bibnamefont{and}
  \bibinfo{author}{\bibfnamefont{J.~E.} \bibnamefont{Butler}},
  \bibinfo{journal}{Appl. Phys. Lett.} \textbf{\bibinfo{volume}{90}},
  \bibinfo{eid}{081110} (\bibinfo{year}{2007}{\natexlab{a}}).

\bibitem[{\citenamefont{Wang et~al.}(2007{\natexlab{b}})\citenamefont{Wang,
  Hanson, Awschalom, Hu, Feygelson, Yang, and Butler}}]{ref:wang2007fct}
\bibinfo{author}{\bibfnamefont{C.~F.} \bibnamefont{Wang}},
  \bibinfo{author}{\bibfnamefont{R.}~\bibnamefont{Hanson}},
  \bibinfo{author}{\bibfnamefont{D.~D.} \bibnamefont{Awschalom}},
  \bibinfo{author}{\bibfnamefont{E.~L.} \bibnamefont{Hu}},
  \bibinfo{author}{\bibfnamefont{T.}~\bibnamefont{Feygelson}},
  \bibinfo{author}{\bibfnamefont{J.}~\bibnamefont{Yang}}, \bibnamefont{and}
  \bibinfo{author}{\bibfnamefont{J.~E.} \bibnamefont{Butler}},
  \bibinfo{journal}{Appl. Phys. Lett.} \textbf{\bibinfo{volume}{91}},
  \bibinfo{pages}{201112} (\bibinfo{year}{2007}{\natexlab{b}}).

\bibitem[{\citenamefont{Bayn and Salzman}(2008)}]{ref:bayn2008uhq}
\bibinfo{author}{\bibfnamefont{I.}~\bibnamefont{Bayn}} \bibnamefont{and}
  \bibinfo{author}{\bibfnamefont{J.}~\bibnamefont{Salzman}},
  \bibinfo{journal}{Optics Express} \textbf{\bibinfo{volume}{16}},
  \bibinfo{pages}{4972} (\bibinfo{year}{2008}).

\bibitem[{\citenamefont{Barclay et~al.}(2008)\citenamefont{Barclay, Painter,
  Santori, Fu, and Beausoleil}}]{ref:barclay2008cie}
\bibinfo{author}{\bibfnamefont{P.}~\bibnamefont{Barclay}},
  \bibinfo{author}{\bibfnamefont{O.}~\bibnamefont{Painter}},
  \bibinfo{author}{\bibfnamefont{C.}~\bibnamefont{Santori}},
  \bibinfo{author}{\bibfnamefont{K.-M.} \bibnamefont{Fu}}, \bibnamefont{and}
  \bibinfo{author}{\bibfnamefont{R.}~\bibnamefont{Beausoleil}},
  \bibinfo{journal}{submitted for publication}  (\bibinfo{year}{2008}).

\bibitem[{\citenamefont{Davies and Hamer}(1976)}]{ref:davies1976oso}
\bibinfo{author}{\bibfnamefont{G.}~\bibnamefont{Davies}} \bibnamefont{and}
  \bibinfo{author}{\bibfnamefont{M.~F.} \bibnamefont{Hamer}},
  \bibinfo{journal}{Proc. R. Soc. London Ser. A}
  \textbf{\bibinfo{volume}{348}}, \bibinfo{pages}{285} (\bibinfo{year}{1976}).

\bibitem[{\citenamefont{Davies et~al.}(1992)\citenamefont{Davies, Lawson,
  Collins, Mainwood, and Sharp}}]{ref:davies1992vrc}
\bibinfo{author}{\bibfnamefont{G.}~\bibnamefont{Davies}},
  \bibinfo{author}{\bibfnamefont{S.~C.} \bibnamefont{Lawson}},
  \bibinfo{author}{\bibfnamefont{A.~T.} \bibnamefont{Collins}},
  \bibinfo{author}{\bibfnamefont{A.}~\bibnamefont{Mainwood}}, \bibnamefont{and}
  \bibinfo{author}{\bibfnamefont{S.~J.} \bibnamefont{Sharp}},
  \bibinfo{journal}{Phys. Rev. B} \textbf{\bibinfo{volume}{46}},
  \bibinfo{pages}{13157} (\bibinfo{year}{1992}).

\bibitem[{\citenamefont{Mita}(1996)}]{ref:mita1996cas}
\bibinfo{author}{\bibfnamefont{Y.}~\bibnamefont{Mita}}, \bibinfo{journal}{Phys.
  Rev. B} \textbf{\bibinfo{volume}{53}}, \bibinfo{pages}{11360}
  (\bibinfo{year}{1996}).

\bibitem[{\citenamefont{Martin et~al.}(1999)\citenamefont{Martin, Wannemacher,
  Teichert, Bischoff, and K{\"o}hler}}]{ref:martin1999gad}
\bibinfo{author}{\bibfnamefont{J.}~\bibnamefont{Martin}},
  \bibinfo{author}{\bibfnamefont{R.}~\bibnamefont{Wannemacher}},
  \bibinfo{author}{\bibfnamefont{J.}~\bibnamefont{Teichert}},
  \bibinfo{author}{\bibfnamefont{L.}~\bibnamefont{Bischoff}}, \bibnamefont{and}
  \bibinfo{author}{\bibfnamefont{B.}~\bibnamefont{K{\"o}hler}},
  \bibinfo{journal}{Appl. Phys. Lett.} \textbf{\bibinfo{volume}{75}},
  \bibinfo{pages}{3096} (\bibinfo{year}{1999}).

\bibitem[{\citenamefont{Waldermann et~al.}(2007)\citenamefont{Waldermann,
  Olivero, Nunn, Surmacz, Wang, Jaksch, Taylor, Walmsley, Draganski, Reichart,
  Greentree, Jamieson, and Prawer}}]{ref:waldermann2007cdc}
\bibinfo{author}{\bibfnamefont{F.}~\bibnamefont{Waldermann}},
  \bibinfo{author}{\bibfnamefont{P.}~\bibnamefont{Olivero}},
  \bibinfo{author}{\bibfnamefont{J.}~\bibnamefont{Nunn}},
  \bibinfo{author}{\bibfnamefont{K.}~\bibnamefont{Surmacz}},
  \bibinfo{author}{\bibfnamefont{Z.}~\bibnamefont{Wang}},
  \bibinfo{author}{\bibfnamefont{D.}~\bibnamefont{Jaksch}},
  \bibinfo{author}{\bibfnamefont{R.}~\bibnamefont{Taylor}},
  \bibinfo{author}{\bibfnamefont{I.}~\bibnamefont{Walmsley}},
  \bibinfo{author}{\bibfnamefont{M.}~\bibnamefont{Draganski}},
  \bibinfo{author}{\bibfnamefont{P.}~\bibnamefont{Reichart}},
  \bibinfo{author}{\bibfnamefont{A.}~\bibnamefont{Greentree}},
  \bibinfo{author}{\bibfnamefont{D.}~\bibnamefont{Jamieson}}, \bibnamefont{and}
  \bibinfo{author}{\bibfnamefont{S.}~\bibnamefont{Prawer}},
  \bibinfo{journal}{Diamond \& Related Materials}
  \textbf{\bibinfo{volume}{16}}, \bibinfo{pages}{1887} (\bibinfo{year}{2007}).

\bibitem[{\citenamefont{Wee et~al.}(2007)\citenamefont{Wee, Tzeng, Han, Chang,
  Fann, Hsu, Chen, and Yu}}]{ref:wee2007tpe}
\bibinfo{author}{\bibfnamefont{T.}~\bibnamefont{Wee}},
  \bibinfo{author}{\bibfnamefont{Y.}~\bibnamefont{Tzeng}},
  \bibinfo{author}{\bibfnamefont{C.}~\bibnamefont{Han}},
  \bibinfo{author}{\bibfnamefont{H.}~\bibnamefont{Chang}},
  \bibinfo{author}{\bibfnamefont{W.}~\bibnamefont{Fann}},
  \bibinfo{author}{\bibfnamefont{J.}~\bibnamefont{Hsu}},
  \bibinfo{author}{\bibfnamefont{K.}~\bibnamefont{Chen}}, \bibnamefont{and}
  \bibinfo{author}{\bibfnamefont{Y.}~\bibnamefont{Yu}}, \bibinfo{journal}{J.
  Phys. Chem. A} \textbf{\bibinfo{volume}{111}}, \bibinfo{pages}{9379}
  (\bibinfo{year}{2007}).

\bibitem[{\citenamefont{Burchard et~al.}(2005)\citenamefont{Burchard, Meijer,
  Popa, Gaebel, Domhan, Wittmann, Jelezko, and
  Wrachtrup}}]{ref:burchard2005gsc}
\bibinfo{author}{\bibfnamefont{B.}~\bibnamefont{Burchard}},
  \bibinfo{author}{\bibfnamefont{J.}~\bibnamefont{Meijer}},
  \bibinfo{author}{\bibfnamefont{I.}~\bibnamefont{Popa}},
  \bibinfo{author}{\bibfnamefont{T.}~\bibnamefont{Gaebel}},
  \bibinfo{author}{\bibfnamefont{M.}~\bibnamefont{Domhan}},
  \bibinfo{author}{\bibfnamefont{C.}~\bibnamefont{Wittmann}},
  \bibinfo{author}{\bibfnamefont{F.}~\bibnamefont{Jelezko}}, \bibnamefont{and}
  \bibinfo{author}{\bibfnamefont{J.}~\bibnamefont{Wrachtrup}},
  \bibinfo{journal}{Appl. Phys. Lett.} \textbf{\bibinfo{volume}{87}},
  \bibinfo{pages}{261909} (\bibinfo{year}{2005}).

\bibitem[{\citenamefont{Rabeau et~al.}(2006)\citenamefont{Rabeau, Reichart,
  Tamanyan, Jamieson, Prawer, Jelezko, Gaebel, Popa, Domhan, and
  Wrachtrup}}]{ref:rabeau2006ils}
\bibinfo{author}{\bibfnamefont{J.}~\bibnamefont{Rabeau}},
  \bibinfo{author}{\bibfnamefont{P.}~\bibnamefont{Reichart}},
  \bibinfo{author}{\bibfnamefont{G.}~\bibnamefont{Tamanyan}},
  \bibinfo{author}{\bibfnamefont{D.}~\bibnamefont{Jamieson}},
  \bibinfo{author}{\bibfnamefont{S.}~\bibnamefont{Prawer}},
  \bibinfo{author}{\bibfnamefont{F.}~\bibnamefont{Jelezko}},
  \bibinfo{author}{\bibfnamefont{T.}~\bibnamefont{Gaebel}},
  \bibinfo{author}{\bibfnamefont{I.}~\bibnamefont{Popa}},
  \bibinfo{author}{\bibfnamefont{M.}~\bibnamefont{Domhan}}, \bibnamefont{and}
  \bibinfo{author}{\bibfnamefont{J.}~\bibnamefont{Wrachtrup}},
  \bibinfo{journal}{Appl. Phys. Lett.} \textbf{\bibinfo{volume}{88}},
  \bibinfo{pages}{023113} (\bibinfo{year}{2006}).

\bibitem[{\citenamefont{Weis et~al.}(2008)\citenamefont{Weis, Schuh, Batra,
  Persaud, Rangelow, Bokor, Lo, Cabrini, Sideras-Haddad, Fuchs, Hanson,
  Awschalom, and Schenkel}}]{ref:weis2008sad}
\bibinfo{author}{\bibfnamefont{C.}~\bibnamefont{Weis}},
  \bibinfo{author}{\bibfnamefont{A.}~\bibnamefont{Schuh}},
  \bibinfo{author}{\bibfnamefont{A.}~\bibnamefont{Batra}},
  \bibinfo{author}{\bibfnamefont{A.}~\bibnamefont{Persaud}},
  \bibinfo{author}{\bibfnamefont{I.}~\bibnamefont{Rangelow}},
  \bibinfo{author}{\bibfnamefont{J.}~\bibnamefont{Bokor}},
  \bibinfo{author}{\bibfnamefont{C.}~\bibnamefont{Lo}},
  \bibinfo{author}{\bibfnamefont{S.}~\bibnamefont{Cabrini}},
  \bibinfo{author}{\bibfnamefont{E.}~\bibnamefont{Sideras-Haddad}},
  \bibinfo{author}{\bibfnamefont{G.}~\bibnamefont{Fuchs}},
  \bibinfo{author}{\bibfnamefont{R.}~\bibnamefont{Hanson}},
  \bibinfo{author}{\bibfnamefont{D.}~\bibnamefont{Awschalom}},
  \bibnamefont{and} \bibinfo{author}{\bibfnamefont{T.}~\bibnamefont{Schenkel}},
  \bibinfo{journal}{J. Vac. Sci Technol. B} \textbf{\bibinfo{volume}{26}},
  \bibinfo{pages}{2596} (\bibinfo{year}{2008}).

\bibitem[{\citenamefont{Allers et~al.}(1998)\citenamefont{Allers, Collins, and
  Hiscock}}]{ref:allers1998air}
\bibinfo{author}{\bibfnamefont{L.}~\bibnamefont{Allers}},
  \bibinfo{author}{\bibfnamefont{A.}~\bibnamefont{Collins}}, \bibnamefont{and}
  \bibinfo{author}{\bibfnamefont{J.}~\bibnamefont{Hiscock}},
  \bibinfo{journal}{Diamond \& Rel. Mat.} \textbf{\bibinfo{volume}{7}},
  \bibinfo{pages}{228} (\bibinfo{year}{1998}).

\bibitem[{\citenamefont{Newton et~al.}(2002)\citenamefont{Newton, Campbell,
  Twitchen, Baker, and Anthony}}]{ref:newton2002red}
\bibinfo{author}{\bibfnamefont{M.}~\bibnamefont{Newton}},
  \bibinfo{author}{\bibfnamefont{B.}~\bibnamefont{Campbell}},
  \bibinfo{author}{\bibfnamefont{D.}~\bibnamefont{Twitchen}},
  \bibinfo{author}{\bibfnamefont{J.}~\bibnamefont{Baker}}, \bibnamefont{and}
  \bibinfo{author}{\bibfnamefont{T.}~\bibnamefont{Anthony}},
  \bibinfo{journal}{Diamond \& Rel. Mat.} \textbf{\bibinfo{volume}{11}},
  \bibinfo{pages}{618} (\bibinfo{year}{2002}).

\bibitem[{\citenamefont{Hounsome et~al.}(2006)\citenamefont{Hounsome, Jones,
  Martineau, Fisher, Shaw, Briddon, and {\"O}berg}}]{ref:hounsome2006obc}
\bibinfo{author}{\bibfnamefont{L.}~\bibnamefont{Hounsome}},
  \bibinfo{author}{\bibfnamefont{R.}~\bibnamefont{Jones}},
  \bibinfo{author}{\bibfnamefont{P.}~\bibnamefont{Martineau}},
  \bibinfo{author}{\bibfnamefont{D.}~\bibnamefont{Fisher}},
  \bibinfo{author}{\bibfnamefont{M.}~\bibnamefont{Shaw}},
  \bibinfo{author}{\bibfnamefont{P.}~\bibnamefont{Briddon}}, \bibnamefont{and}
  \bibinfo{author}{\bibfnamefont{S.}~\bibnamefont{{\"O}berg}},
  \bibinfo{journal}{Phys. Rev. B} \textbf{\bibinfo{volume}{73}},
  \bibinfo{pages}{125203} (\bibinfo{year}{2006}).

\bibitem[{\citenamefont{Nelson et~al.}(1983)\citenamefont{Nelson, Hudson,
  Mazey, and Piller}}]{ref:nelson1983dsi}
\bibinfo{author}{\bibfnamefont{R.}~\bibnamefont{Nelson}},
  \bibinfo{author}{\bibfnamefont{J.}~\bibnamefont{Hudson}},
  \bibinfo{author}{\bibfnamefont{D.}~\bibnamefont{Mazey}}, \bibnamefont{and}
  \bibinfo{author}{\bibfnamefont{R.}~\bibnamefont{Piller}},
  \bibinfo{journal}{Proceedings of the Royal Society of London. Series A,
  Mathematical and Physical Sciences (1934-1990)}
  \textbf{\bibinfo{volume}{386}}, \bibinfo{pages}{211} (\bibinfo{year}{1983}).

\bibitem[{\citenamefont{Gippius et~al.}(1999)\citenamefont{Gippius,
  Khmelnitskiy, Dravin, and Tkachenko}}]{ref:gippius1999fcg}
\bibinfo{author}{\bibfnamefont{A.}~\bibnamefont{Gippius}},
  \bibinfo{author}{\bibfnamefont{R.}~\bibnamefont{Khmelnitskiy}},
  \bibinfo{author}{\bibfnamefont{V.}~\bibnamefont{Dravin}}, \bibnamefont{and}
  \bibinfo{author}{\bibfnamefont{S.}~\bibnamefont{Tkachenko}},
  \bibinfo{journal}{Diamond \& Related Materials} \textbf{\bibinfo{volume}{8}},
  \bibinfo{pages}{1631} (\bibinfo{year}{1999}).

\bibitem[{\citenamefont{Zeigler}(2008)}]{ref:zeigler2008sri}
\bibinfo{author}{\bibfnamefont{J.}~\bibnamefont{Zeigler}},
  \emph{\bibinfo{title}{The stopping range of ions in matter, srim-2008}}
  (\bibinfo{year}{2008}).

\bibitem[{\citenamefont{Santori et~al.}(2006)\citenamefont{Santori, Fattal,
  Spillane, Fiorentino, Beausoleil, Greentree, Olivero, Draganski, Rabeau,
  Reichart, Gibson, Rubanov, Jamieson, and Prawer}}]{ref:santori2006cpta}
\bibinfo{author}{\bibfnamefont{C.}~\bibnamefont{Santori}},
  \bibinfo{author}{\bibfnamefont{D.}~\bibnamefont{Fattal}},
  \bibinfo{author}{\bibfnamefont{S.}~\bibnamefont{Spillane}},
  \bibinfo{author}{\bibfnamefont{M.}~\bibnamefont{Fiorentino}},
  \bibinfo{author}{\bibfnamefont{R.}~\bibnamefont{Beausoleil}},
  \bibinfo{author}{\bibfnamefont{A.}~\bibnamefont{Greentree}},
  \bibinfo{author}{\bibfnamefont{P.}~\bibnamefont{Olivero}},
  \bibinfo{author}{\bibfnamefont{M.}~\bibnamefont{Draganski}},
  \bibinfo{author}{\bibfnamefont{J.}~\bibnamefont{Rabeau}},
  \bibinfo{author}{\bibfnamefont{P.}~\bibnamefont{Reichart}},
  \bibinfo{author}{\bibfnamefont{B.}~\bibnamefont{Gibson}},
  \bibinfo{author}{\bibfnamefont{S.}~\bibnamefont{Rubanov}},
  \bibinfo{author}{\bibfnamefont{D.}~\bibnamefont{Jamieson}}, \bibnamefont{and}
  \bibinfo{author}{\bibfnamefont{S.}~\bibnamefont{Prawer}},
  \bibinfo{journal}{Optics Express} \textbf{\bibinfo{volume}{14}},
  \bibinfo{pages}{7986} (\bibinfo{year}{2006}).

\bibitem[{\citenamefont{Leech et~al.}(2001)\citenamefont{Leech, Reeves, and
  Holland}}]{ref:leech2001jms}
\bibinfo{author}{\bibfnamefont{P.~W.} \bibnamefont{Leech}},
  \bibinfo{author}{\bibfnamefont{G.~K.} \bibnamefont{Reeves}},
  \bibnamefont{and} \bibinfo{author}{\bibfnamefont{A.}~\bibnamefont{Holland}},
  \bibinfo{journal}{J. Mat. Sci.} \textbf{\bibinfo{volume}{36}},
  \bibinfo{pages}{3453} (\bibinfo{year}{2001}).

\bibitem[{\citenamefont{Lukosz and Kunz}(1977)}]{ref:lukosz1977leb}
\bibinfo{author}{\bibfnamefont{W.}~\bibnamefont{Lukosz}} \bibnamefont{and}
  \bibinfo{author}{\bibfnamefont{R.}~\bibnamefont{Kunz}}, \bibinfo{journal}{J.
  Opt. Soc. Am.} \textbf{\bibinfo{volume}{67}}, \bibinfo{pages}{1607}
  (\bibinfo{year}{1977}).

\bibitem[{\citenamefont{Liu et~al.}(2001)\citenamefont{Liu, Lin, Huang, Guo,
  and Duan}}]{ref:liu2001sea}
\bibinfo{author}{\bibfnamefont{Y.}~\bibnamefont{Liu}},
  \bibinfo{author}{\bibfnamefont{J.}~\bibnamefont{Lin}},
  \bibinfo{author}{\bibfnamefont{G.}~\bibnamefont{Huang}},
  \bibinfo{author}{\bibfnamefont{Y.}~\bibnamefont{Guo}}, \bibnamefont{and}
  \bibinfo{author}{\bibfnamefont{C.}~\bibnamefont{Duan}}, \bibinfo{journal}{J.
  Opt. Soc. Am. B} \textbf{\bibinfo{volume}{18}}, \bibinfo{pages}{666}
  (\bibinfo{year}{2001}).

\bibitem[{\citenamefont{{Aharonovich et al.}}(2009)}]{ref:aharonovich2008}
\bibinfo{author}{\bibfnamefont{I.}~\bibnamefont{{Aharonovich et al.}}},
  \bibinfo{journal}{to be published}  (\bibinfo{year}{2009}).

\bibitem[{\citenamefont{Iakoubovskii et~al.}(2000)\citenamefont{Iakoubovskii,
  Adriaenssens, and Nesladek}}]{ref:iakoubovskii2000pvr}
\bibinfo{author}{\bibfnamefont{K.}~\bibnamefont{Iakoubovskii}},
  \bibinfo{author}{\bibfnamefont{G.}~\bibnamefont{Adriaenssens}},
  \bibnamefont{and} \bibinfo{author}{\bibfnamefont{M.}~\bibnamefont{Nesladek}},
  \bibinfo{journal}{J. Phys. Cond. Mat.} \textbf{\bibinfo{volume}{12}},
  \bibinfo{pages}{189} (\bibinfo{year}{2000}).

\bibitem[{\citenamefont{Wotherspoon et~al.}(2003)\citenamefont{Wotherspoon,
  Steeds, Catmull, and Butler}}]{ref:wotherspoon2003pap}
\bibinfo{author}{\bibfnamefont{A.}~\bibnamefont{Wotherspoon}},
  \bibinfo{author}{\bibfnamefont{J.}~\bibnamefont{Steeds}},
  \bibinfo{author}{\bibfnamefont{B.}~\bibnamefont{Catmull}}, \bibnamefont{and}
  \bibinfo{author}{\bibfnamefont{J.}~\bibnamefont{Butler}},
  \bibinfo{journal}{Diamond \& Rel. Mat.} \textbf{\bibinfo{volume}{12}},
  \bibinfo{pages}{652} (\bibinfo{year}{2003}).

\bibitem[{\citenamefont{Manson and Harrison}(2005)}]{ref:manson2005pin}
\bibinfo{author}{\bibfnamefont{N.}~\bibnamefont{Manson}} \bibnamefont{and}
  \bibinfo{author}{\bibfnamefont{J.}~\bibnamefont{Harrison}},
  \bibinfo{journal}{Diamond \& Rel. Mat.} \textbf{\bibinfo{volume}{14}},
  \bibinfo{pages}{1705} (\bibinfo{year}{2005}).

\bibitem[{\citenamefont{Farrer}(1969)}]{ref:farrer1969snd}
\bibinfo{author}{\bibfnamefont{R.}~\bibnamefont{Farrer}},
  \bibinfo{journal}{Solid State Comm.} \textbf{\bibinfo{volume}{7}},
  \bibinfo{pages}{685} (\bibinfo{year}{1969}).

\bibitem[{\citenamefont{Steeds et~al.}(2000)\citenamefont{Steeds, Charles,
  Davies, and Griffin}}]{ref:steeds2000pmt}
\bibinfo{author}{\bibfnamefont{J.}~\bibnamefont{Steeds}},
  \bibinfo{author}{\bibfnamefont{S.}~\bibnamefont{Charles}},
  \bibinfo{author}{\bibfnamefont{J.}~\bibnamefont{Davies}}, \bibnamefont{and}
  \bibinfo{author}{\bibfnamefont{I.}~\bibnamefont{Griffin}},
  \bibinfo{journal}{Diamond \& Rel. Mat.} \textbf{\bibinfo{volume}{9}},
  \bibinfo{pages}{397} (\bibinfo{year}{2000}).

\bibitem[{\citenamefont{Goss et~al.}(2005)\citenamefont{Goss, Briddon, Rayson,
  Sque, and Jones}}]{ref:goss2005vic}
\bibinfo{author}{\bibfnamefont{J.}~\bibnamefont{Goss}},
  \bibinfo{author}{\bibfnamefont{P.}~\bibnamefont{Briddon}},
  \bibinfo{author}{\bibfnamefont{M.}~\bibnamefont{Rayson}},
  \bibinfo{author}{\bibfnamefont{S.}~\bibnamefont{Sque}}, \bibnamefont{and}
  \bibinfo{author}{\bibfnamefont{R.}~\bibnamefont{Jones}},
  \bibinfo{journal}{Phys. Rev. B} \textbf{\bibinfo{volume}{72}},
  \bibinfo{pages}{35214} (\bibinfo{year}{2005}).

\bibitem[{\citenamefont{Ristein}(2000)}]{ref:ristein2000epd}
\bibinfo{author}{\bibfnamefont{J.}~\bibnamefont{Ristein}},
  \bibinfo{journal}{Diamond \& Rel. Mat.} \textbf{\bibinfo{volume}{9}},
  \bibinfo{pages}{1129} (\bibinfo{year}{2000}).

\bibitem[{\citenamefont{Dannefaer et~al.}(2001)\citenamefont{Dannefaer, Pu, and
  Kerr}}]{ref:dannefaer2001pas}
\bibinfo{author}{\bibfnamefont{S.}~\bibnamefont{Dannefaer}},
  \bibinfo{author}{\bibfnamefont{A.}~\bibnamefont{Pu}}, \bibnamefont{and}
  \bibinfo{author}{\bibfnamefont{D.}~\bibnamefont{Kerr}},
  \bibinfo{journal}{Diamond \& Rel. Mat.} \textbf{\bibinfo{volume}{10}},
  \bibinfo{pages}{2113} (\bibinfo{year}{2001}).

\end{thebibliography}
\end{document}